\documentclass[fleqn,usenatbib]{mnras}

\usepackage{newtxtext,newtxmath}

\usepackage[T1]{fontenc}
\usepackage{blindtext}
\usepackage{float}

\DeclareRobustCommand{\VAN}[3]{#2}
\let\VANthebibliography\thebibliography
\def\thebibliography{\DeclareRobustCommand{\VAN}[3]{##3}\VANthebibliography}

\usepackage{graphicx}	
\usepackage{amsmath}	
\usepackage{stfloats}
\usepackage{tipa}

\defcitealias{ellisonLowRedshiftPoststarburst2025a}{E25}

\title[EMBERS I]{EMBERS I: Low redshift post-starburst galaxies are frequently  depleted in molecular gas relative to star forming progenitors}

\author[B. F. Rasmussen et al.]{Ben F. Rasmussen$^{1}$\thanks{E-mail: benfrasmussen@uvic.ca},
María Jesús Jiménez-Donaire$^{2,3}$,
Sara L. Ellison$^{1}$,
Vivienne Wild$^{4}$,
\newauthor Kate Rowlands$^{2,5}$,
Qifeng Huang$^{6,7}$,
Jing Wang$^{6}$,
Dong Yang$^{6,7}$,
Scott Wilkinson$^{1}$,
Blake Ledger$^{1}$,
\newauthor
Toby Brown$^{8}$,
Ho-Hin Leung$^{9}$,
Shoshannah Byrne-Mamahit$^{1}$
\vspace{2mm}
\\
$^{1}$Department of Physics \& Astronomy, University of Victoria, Finnerty Road, Victoria, BC V8P 1A1, Canada\\
$^{2}$AURA for ESA, Space Telescope Science Institute, 3700 San Martin Drive, Baltimore, MD 21218, USA\\
$^{3}$Observatorio Astronómico Nacional (IGN), C/Alfonso XII 3, 28014, Madrid, Spain\\
$^{4}$School of Physics and Astronomy, University of St Andrews, North Haugh, St Andrews, KY16 9SS, U.K.\\
$^{5}$William H. Miller III Department of Physics and Astronomy, Johns Hopkins University, Baltimore, MD 21218, USA\\
$^{6}$Kavli Institute for Astronomy and Astrophysics, Peking University, Beijing 100871, China \\
$^{7}$Department of Astronomy, School of Physics, Peking University, Beijing 100871, China\\
$^{8}$National Research Council of Canada, Herzberg Astronomy and Astrophysics Research Centre, 5071 W. Saanich Rd. Victoria, BC, V9E
2E7, Canada\\
$^{9}$SUPA, Institute for Astronomy, University of Edinburgh, Royal Observatory, Edinburgh EH9 3HJ, UK
}

\date{Accepted XXX. Received YYY; in original form ZZZ}

\pubyear{\the\year{}}

\begin{document}
\label{firstpage}
\pagerange{\pageref{firstpage}--\pageref{lastpage}}
\maketitle

\begin{abstract}
The cold gas content of post-starburst galaxies (PSBs) provides important insight into the mechanisms that drive rapid quenching, but a multiphase assessment of both the atomic and molecular gas in PSBs does not yet exist. We introduce the Ensemble of Multiphase Baryons Evolving in Rapidly-quenching Systems, or EMBERS, a homogeneously selected, nearly mass- and redshift-complete survey of the global atomic ($\text{H\sc{i}}$) and molecular gas (H$_2$) in PSBs, observed with the Five Hundred-metre Aperture Spherical Telescope (FAST) and the Institut de radioastronomie millimétrique (IRAM) 30m telescope. We present new CO(1--0) observations for 52 PSBs with the IRAM 30m, which, combined with 9 archival observations, gives a total H$_2$ sample of 61, of which 58/61 have ancillary H$\text{\sc{i}}$ measurements. We detect CO(1--0) in 34/61 galaxies, corresponding to molecular gas fractions ($f_{\rm H_2} = \rm M_{H_2}/M_{\ast}$) ranging from two to 250 per cent. By comparing with a stellar-mass matched star-forming (SF) control sample from xCOLD GASS, we find that PSBs on \textit{average} are $\sim 0.3-0.6$ dex depleted in H$_2$. However, considering both $\text{H\sc{i}}$ and H$_2$, \textit{individual} PSBs host diverse gas reservoirs ranging from gas-rich in both phases, elevated in one phase, or gas-poor, the latter of which is common at lower stellar mass. The existence of gas-normal and gas-depleted PSBs in both phases suggests that some PSBs may rejuvenate their star formation, but the rapid shutdown of star formation in others is likely terminal. Despite this diversity, the majority of EMBERS PSBs are gas-poor compared to SF controls, with the typical PSB hosting gas reservoirs intermediate to those found in star-forming and quenched galaxies.

\end{abstract}


\section{Introduction}

Observational studies of large samples of massive galaxies have shown that the majority exist in one of two regimes: actively star-forming (SF), blue, and disk-dominated spirals, or quiescent, red ellipticals \citep{strateva_2001,baldry_2004,driver_2006,brammer_2009,wuyts_2011}. Photometric and spectroscopic surveys highlight this bimodality further when we consider the star formation rates (SFRs) as a function of stellar mass \citep{kauffmann_stellar_2003,brinchmann_2004,salim_2007}. Star-forming galaxies form a tight \textit{main sequence}, while quenched galaxies populate a separate distribution lying an order of magnitude lower in SFR than the main sequence at fixed stellar mass \citep{noeske_2007,speagle_2014,bluck_agn_2020}. This bimodality in properties extends to cold gas content, where SF galaxies are generally gas-rich, whereas quenched galaxies are typically gas-poor \citep{saintongeXCOLDGASSComplete2017a, catinella_xgass_2018, colombo_2020}. 

The star formation rate of a galaxy is strongly correlated with gas content, as the cold gas in the interstellar medium (ISM) is the primary fuel for new stars \citep{kennicutt_1998,krumholz_2012, kennicutt_2012, schinnerer_leroy_2024rev}. The ISM consists of multiple phases, with each phase becoming progressively more fundamental to star formation as the density of the phase increases \citep{saintaonge_catinella_annurev_2022}. The atomic gas phase, traced by 21 cm emission from $\text{H\sc{i}}$, is diffuse and is present in extended reservoirs \citep{walter_2008, putman_2012}. Whereas the molecular gas phase, made up of mostly $\rm H_2$ but traced by the CO molecule, exists in denser regions preferentially in the galactic disk \citep{bigiel_2008,bolatto_alpha_2013, leroy_2021}. The molecular gas reservoir is created when atomic gas cools, and so an understanding of both gas phases is required to fully capture the fueling of star formation in a galaxy \citep{krumholz_2009, saintongeMolecularAtomicGas2016}. Disruptions to either the atomic or molecular gas (or both) are necessary if a galaxy were to transition from blue to red. 

The rising fraction of quenched galaxies over cosmic time \textit{necessitates} a transition from blue and active to red and dead \citep{hopkins_2007,ilbert_2013,muzzin_2013}. The mechanisms responsible and the relative importance of them for the migration of galaxies off the main sequence to the quenched population remain an open question in the literature (e.g., \citeauthor{mann_2018} \citeyear{mann_2018}; \citeauthor{ellison_2021} \citeyear{ellison_2021}; \citeauthor{piotrowska_2022} \citeyear{piotrowska_2022}; \citeauthor{brown_2023} \citeyear{brown_2023}). Possible processes responsible include gas removal due to ram pressure stripping \citep{woo_2013, poggianti_2017, bosselli_2022}, active galactic nuclei (AGN) feedback \citep{terrazas_2016,chen_2020,bluck_2020}, consumption or inefficient refueling of a galaxy's gas reservoir \citep{peng_2015, whitaker_2021}, an inability to form gas of sufficient densities needed for star formation due to turbulence of the gas inside the galaxy \citep{gensior_2020, smercinaFallResolvingMolecular2022}, and major mergers that disrupt the ISM \citep{belli_2019, ellison_2022,ellison_psbs_pm_2024, sargentMolecularGasContent2025}.

To disentangle the physical processes that result in quenching, it is useful to study galaxies that are in the process of shutting down their star formation. Galaxies caught in the act of quenching, or post-starburst (PSB) galaxies, are selected to have recently formed stars in a starburst event but show no signatures of ongoing star formation \citep{dressler_gunn_1983,couch_sharples_1987,goto266E+AGalaxies2005,wild_2007}. The selection criteria for how PSBs are identified differ (see \citeauthor{frenchEvolutionPoststarburstPhase2021a} \citeyear{frenchEvolutionPoststarburstPhase2021a} for a summary of PSBs and methods), but signatures of a recently formed stellar population combined with a lack of \textit{very} young stars are necessary to find galaxies in the process of quenching. PSBs are rare, making up $<1\%$ of galaxies, but can provide considerable insight into the mechanisms behind rapid galaxy evolution \citep{wild_2009, pawlik_2016, alataloWelcomeTwilightZone2017, rowlands_2018, smercinaFallDustGas2018,bezanson_2022}.

Recent work has shown that the majority of PSBs show morphological signatures of merging with another galaxy, but that not all PSBs are mergers, suggesting that this mechanism alone is insufficient to explain the rapid shutdown of star formation \citep{pawlikOriginsPoststarburstGalaxies2018,sazonova_2021,wilkinson_2022,ellison_psbs_pm_2024}. Similarly, some PSBs have been shown to host active galactic nuclei, but that it is not the smoking gun needed to explain the entire population of PSBs \citep{ellison_2022, lanz_2022}. There is a delay in timescales between the quenching seen in PSBs and the onset of AGN activity, and the presence of AGN is not ubiquitous across all PSBs \citep{wild_2010, yesuf_2014,yesuf_2020,krishna_2025}. Although both AGN and mergers seem to play a role in the rapid shutdown of star formation seen in PSBs, it is their effect on the gas reservoirs that must ultimately result in quenching, and so it is necessary to study the gas contents of rapidly quenching galaxies. 

The global molecular gas in PSBs has been studied in a number of previous works with samples ranging from 15 to over 100 \citep{frenchDISCOVERYLARGEMOLECULAR2015g,rowlands_psbs_2015,alataloSHOCKEDPOSTSTARBURSTGALAXY2016a,yesuf_2017,baron_2023}. The consensus of most studies is that PSBs host considerable molecular gas reservoirs with gas fractions ranging from those typical of active star-forming galaxies to the gas-rich end of the quiescent galaxy population. To reconcile the large gas reservoirs with the shutdown of star formation, \cite{otterResolvedMolecularGas2022} and \cite{smercinaFallResolvingMolecular2022} obtained {resolved} observations of molecular gas in PSBs and found that the star formation suppression seen is potentially due to highly compact, turbulent gas reservoirs with inefficient star formation. The inability to form sufficient densities conducive for star formation due to turbulence is supported by low fractions of dense gas, traced by the HCN molecule, present in PSBs measured by \cite{french_2018} and \cite{frenchStateMolecularGas2023c}. The picture of rapid quenching and the role of molecular gas is becoming clearer, but previous works have lacked a uniform selection criterion and observation sensitivity threshold, without a suitable sample of normal galaxies to compare with. Additionally, the interplay between the atomic and molecular gas phases has not yet been studied as there has been far less work done on the atomic gas phase.

H$\text{\sc{i}}$ observations exist in a handful of previous works obtaining measurements for PSB sample sizes ranging from 5 to 15 galaxies, with inhomogeneous selection criteria and observation depth \citep{bravo_alfaro_2001,chang_2001,buyleContentE+AGalaxies2006, zwaanColdGasContent2013,li_2023}. Previous studies of the atomic phase have shown a diversity of atomic gas properties, with the majority of PSBs retaining an atomic gas reservoir throughout the post-starburst phase. However, the small sample sizes available have precluded broader conclusions about PSBs as a population.

The sample size limitations, as well as the lack of uniform selection and sensitivity criteria for atomic gas in PSBs were not fully addressed until \cite{ellisonLowRedshiftPoststarburst2025a}, where we present measurements of atomic gas in 68 PSBs with observation sensitivities matched to the xGASS\footnote{Acronyms: xGASS - the extended GALEX Arecibo SDSS Survey; xCOLD GASS - the extended CO Legacy Database for GASS} survey \citep{catinella_2010, catinella_xgass_2018}. In \citet{ellisonLowRedshiftPoststarburst2025a} we find that many of the rapidly quenching galaxies host considerable atomic gas masses, with a mild reduction when explicitly compared to a star-forming progenitor population, and that the wholesale removal of H$\text{\sc{i}}$ gas is not responsible for the rapid truncation of star formation. \citet{huang_2025} analyzed the shape of the H$\text{\sc{i}}$ spectra from \citet{ellisonLowRedshiftPoststarburst2025a} and showed that although there are significant reservoirs of atomic gas, it is preferentially centrally concentrated compared to normal galaxies in xGASS.

The observations from \citet{ellisonLowRedshiftPoststarburst2025a} (hereafter \citetalias{ellisonLowRedshiftPoststarburst2025a}) obtained with the Five hundred meter Aperture Spherical Telescope (FAST) make up the first component of the Ensemble of Multiphase Baryons Evolving in Rapidly-quenching Systems, or EMBERS survey, which we introduce here. We combine the measurements of \citetalias{ellisonLowRedshiftPoststarburst2025a} with newly obtained observations of the molecular phase, traced by CO and taken with the Institut de radioastronomie millimétrique (IRAM) 30m telescope, for the same sample of post-starburst galaxies. EMBERS is thus the first survey able to track the evolution of both the atomic and molecular gas phases in PSBs. The EMBERS sample is explicitly designed to match the observing parameters of the xGASS \citep{catinella_xgass_2018} and xCOLD GASS\footnotemark[1] \citep{saintongeXCOLDGASSComplete2017a} surveys with a large sample of PSBs, selected in a homogeneous, mass and redshift-complete manner. As a result, we can compare both the atomic and molecular gas phases of PSBs directly to normal galaxies in a controlled way for the first time. 

This paper is organized as follows. In Section \ref{sec:methods}, we describe the methods of sample selection and present our new CO(1--0) observations and their reduction. In Section \ref{sec:results}, we present our results and discuss their implications and the impacts of our assumptions in Section \ref{sec:discussion}. Finally, we summarize and conclude our work in Section \ref{sec:conclusions}. Throughout, we adopt a cosmology with
H$_0 = 70\ \mathrm{km\ s^{-1}\ Mpc^{-1}}$, $\Omega_\mathrm{M} = 0.3$, and $\Omega_{\Lambda} = 0.7$.

\section{Data and Methods}
\label{sec:methods}
\subsection{Sample selection}
\label{sec:sample} 

\begin{figure}
\centering
    \includegraphics[scale=0.9]{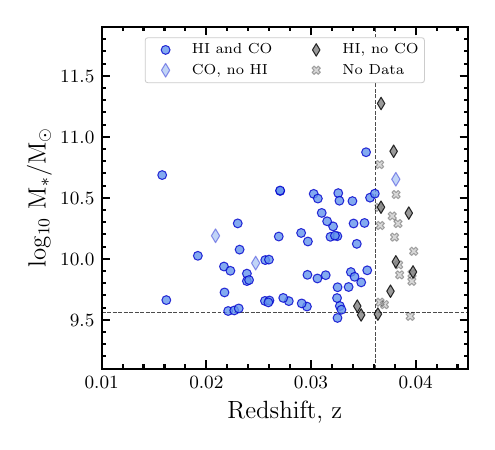}
    \caption{Redshift-stellar mass distribution of all PSBs in EMBERS. Horizontal and vertical dashed lines correspond to the adjusted cuts of $z<0.0362$ and log$_{10}$ M$_{\ast}$/M$_{\odot}$ > 9.56 that define the boundaries of our CO observations. Blue circles have both CO and H$\text{\sc{i}}$ measurements, blue diamonds have only CO, grey diamonds have only H$\text{\sc{i}}$, grey crosses were not observed in either program. One PSB (EMB13) was not observed in the IRAM 30m program but falls within our cuts while two PSBs (EMB20 and 62) were observed with IRAM but the FAST observations were compromised due to a close companion and RFI. {Additionally, EMB73 has an archival CO measurement but was not observed in H$\text{\sc{i}}$.} Our sample is then {nearly} mass and redshift complete within our prescribed boundaries in both H$\text{\sc{i}}$ and CO. }

    \label{fig:mass redshift}
\end{figure}

All EMBERS PSBs are selected from the Sloan Digital Sky Survey (SDSS) DR7, originating from two separate but complementary catalogs, encompassing the diversity of PSB types. Traditional E+A PSBs are selected using classifications\footnote{Classifications and catalogs are available at:
\url{http://www.phys.nthu.edu.tw/\~tomo/cv/index.html} and \url{http://star-www.st-andrews.ac.uk/~vw8/downloads/index.html}}
following the methods of \cite{goto266E+AGalaxies2005} and \cite{goto_2007}. These correspond to equivalent width (EW) cuts (with positive values denoting absorption and negative values denoting emission) of: EW(H$\delta$) $>$ $5.0$ \AA, EW($[\rm{OII}]_{\rm \lambda3727}$) $>$ $-2.5$ \AA, EW(H$\alpha$) $>$ $-3.0$ \AA. The traditional E+A PSB galaxies are combined with PSBs identified through the principal component analysis (PCA) selection of \cite{wild_2007}\footnotemark[2]. This method selects galaxies with an excess of Balmer absorption given the age of the stellar population, ignoring emission line strengths. To reject the selection of dusty, star-forming interlopers, an additional mass-dependent cut in Balmer decrement is made \citep{pawlikOriginsPoststarburstGalaxies2018}. For a complete description of the PCA selection criteria see \cite{wild_2007} and \cite{wilkinson_2022}. 

From this SDSS DR7 derived PSB parent sample, we initially select galaxies with inclusive cuts on mass and redshift of 0.01 $<$ $z$ $<$ 0.04 and log$_{10}$ M$_\ast$/M$_{\odot}$ $>$ 9.5. These cuts yielded a complete and representative sample of 114 PSBs that are suitably observable in multiple gas phases. The SDSS optical spectra were then visually inspected to confirm the presence of strong Balmer absorption features we expect of PSBs, rejecting three. A further 25 galaxies were removed due to contamination in the FAST beam for galaxies with spectroscopic companions (see \citetalias{ellisonLowRedshiftPoststarburst2025a} for more details). {Our final sample consists of 86 PSBs, which includes every galaxy in the SDSS that falls within our selection criteria outlined above.}

The first step in the EMBERS campaign was to search for existing H$\text{\sc{i}}$ observations from ALFALFA \citep{haynes_alfalfa_2018}, xGASS \citep{catinella_xgass_2018}, HI-MaNGA \citep{stark_himanga_2021}, and FASHI \citep{zhang_fashi_2024} finding 26 archival measurements for EMBERS PSBs.\footnote{Acronyms: ALFALFA — Arecibo Legacy Fast Arecibo L-band Feed Array survey; HI-MaNGA — H$\text{\sc{i}}$ follow-up to the Mapping Nearby Galaxies at Apache Point Observatory survey; FASHI — FAST All Sky H$\text{\sc{i}}$ survey.} Additionally, 44 new observations were made by \citetalias{ellisonLowRedshiftPoststarburst2025a} with FAST resulting in an H$\text{\sc{i}}$ sample of 70 galaxies. Two of these measurements were removed from the final sample, one due to radio-frequency interference (EMB20) and the other due to a close companion not identified in the first contamination check (EMB62). Both of the removed galaxies are included and noted here for completeness. Table \ref{tab:knowndata} summarizes the entire EMBERS sample as well as the relevant details of the observations from \citetalias{ellisonLowRedshiftPoststarburst2025a}. {Galaxies that were not observed in H$\text{\sc{i}}$ but fit the EMBERS selection criteria are included in Table \ref{tab:knowndata} and Figure \ref{fig:mass redshift} for completeness.}

Of the 70 galaxies with H$\text{\sc{i}}$ observations, we searched for archival global CO measurements that fulfill our goal detection threshold (see next section). In doing so, we found  {five} observations from \citet{frenchDISCOVERYLARGEMOLECULAR2015g} of which two are upper limits, two detections from \citet{alataloSHOCKEDPOSTSTARBURSTGALAXY2016a}, and one detection in the xCOLD GASS survey \citep{saintongeXCOLDGASSComplete2017a} which we add to our PSB sample and remove from the control sample drawn from xCOLD GASS in later sections. One galaxy, EMB73, has an archival CO measurement from \citet{frenchDISCOVERYLARGEMOLECULAR2015g} which we include, despite no H$\text{\sc{i}}$ observation. This leaves us with 61 galaxies with H$\text{\sc{i}}$ that require new CO measurements.

\subsection{Observations and data reduction}
\label{sec:observations}

\begin{table*}
    \caption{Complete EMBERS sample galactic properties, H$\text{\sc{i}}$ information, and identifiers}
    \centering
    \setlength{\tabcolsep}{10pt}
    \begin{tabular}{ccccccccc}
    \hline 
    EMBERS ID & SDSS DR7 objID & RA & DEC  & $z$ &$\log \rm M_*$ & $\log \rm M_{\rm H\text{\sc{i}}}$ & H$\text{\sc{i}}$ Source & {Alt ID} \\
     &       & ($^{\circ}$) & ($^{\circ}$) &  & ($\rm M_{\sun}$) & ($\rm M_{\sun}$) & \\
    \hline
    EMB00 & 587722984440463382 & 216.55409 & 0.86058 & 0.031859 & 10.18 & 9.06 ± 0.05 & 5 & —\\
    EMB01 & 587724197741527049 & 16.57401 & 14.06501 & 0.038471 & 9.87 & — & 0 & —\\
    EMB02 & 587725489988960505 & 257.13931 & 57.47670 & 0.029610 & 9.61 & 8.86 & 3 & —\\
    EMB03 & 587726032256630848 & 198.46830 & 2.13257 & 0.030259 & 10.53 & 9.40 ± 0.04 & 5 & {J1313+0207$^2$}\\
    EMB04 & 587727179012046875 & 59.21853 & -6.17534 & 0.037017 & 9.62 & — & 0 & —\\
    EMB05 & 587728918984392809 & 215.59795 & 61.69753 & 0.036555 & 10.77 & — & 0 & —\\
    EMB06 & 587728931343630410 & 136.47356 & 48.82022 & 0.039342 & 10.37 & 9.69 & 4 & —\\
    EMB07 & 587729408621609096 & 259.53273 & 30.12902 & 0.029660 & 9.87 & 8.67 ± 0.06 & 5 & —\\
    EMB08 & 587729652890206755 & 256.43331 & 31.41377 & 0.034796 & 9.81 & 9.00 ± 0.05 & 5 & —\\
    EMB09 & 587730023333232703 & 230.52173 & 5.85486 & 0.035642 & 10.50 & 9.89 ± 0.05 & 1 & —\\
    EMB10 & 587730773885059094 & 351.22557 & 14.21521 & 0.025628 & 9.99 & 9.45 ± 0.06 & 1 & —\\
    EMB11 & 587730847963545655 & 318.50226 & 0.53510 & 0.026921 & 10.18 & 9.40 ± 0.04 & 5 & {EAH04$^1$}\\
    EMB12 & 587730846889869791 & 318.67133 & -0.41098 & 0.032113 & 10.27 & 8.62 ± 0.06 & 5 & —\\
    EMB13 & 587731174382502294 & 319.95072 & 0.67272 & 0.034424 & 9.61 & 9.08 ± 0.05 & 5 & —\\
    EMB14 & 587731186735186207 & 347.26187 & 0.26689 & 0.032520 & 10.19 & 9.19 ± 0.04 & 5 & —\\
    EMB15 & 587731522273607944 & 120.57663 & 32.53550 & 0.038310 & 10.29 & — & 0 & —\\
    EMB16 & 587731886809808959 & 123.85738 & 37.34052 & 0.039747 & 9.89 & 9.67 & 3 & —\\
    EMB17 & 587731185117954089 & 332.02546 & -0.90695 & 0.037982 & 10.18 & — & 0 & —\\
    EMB18 & 587731512082956346 & 50.88860 & -0.43856 & 0.023876 & 9.88 & <8.72 & 5 & —\\
    EMB19 & 587731512619106324 & 49.22882 & -0.04199 & 0.023179 & 10.08 & 8.60 & 3 & {EAS02$^1$}\\
    EMB20 & 587731681190740132 & 129.18887 & 39.83509 & 0.024715 & 9.96 & 0.00$^\ddagger$ & 5 & —\\
    EMB21 & 587732482747400295 & 176.81117 & 49.18526 & 0.025601 & 9.65 & 8.51 ± 0.09 & 5 & —\\
    EMB22 & 587732580982521898 & 169.78179 & 58.05398 & 0.032604 & 10.54 & 9.13 ± 0.05 & 5 & {EAS07$^1$}\\
    EMB23 & 587732701250519063 & 149.29786 & 5.20132 & 0.021686 & 9.94 & <8.56 & 5 & —\\
    EMB24 & 587734622705811566 & 131.15756 & 32.90645 & 0.031543 & 10.31 & 9.00 ± 0.05 & 5 & —\\
    EMB25 & 587734891673026666 & 165.36402 & 8.42062 & 0.030614 & 9.84 & 9.79 ± 0.06 & 1 & —\\
    EMB26 & 587732771584671859 & 164.59071 & 9.45391 & 0.033594 & 9.77 & 8.98 ± 0.05$^\dagger$ & 5 & —\\
    EMB27 & 587733195160485975 & 196.35758 & 53.59174 & 0.038128 & 10.53 & — & 0 & —\\
    EMB28 & 587735349098971209 & 154.24553 & 13.39929 & 0.032541 & 9.77 & 8.40 ± 0.11 & 5 & —\\
    EMB29 & 587735742620107149 & 248.03348 & 25.65341 & 0.039485 & 9.53 & — & 0 & —\\
    EMB30 & 587735742621286560 & 250.27812 & 23.84094 & 0.036491 & 10.42 & — & 0 & —\\
    EMB31 & 587735347486457882 & 149.92061 & 11.53314 & 0.036687 & 10.42 & 9.40 ± 0.07 & 1 & —\\
    EMB32 & 587735431226130460 & 191.61184 & 50.79205 & 0.027056 & 10.56 & 8.84 ± 0.06$^\dagger$ & 5 & {EAS09$^1$}\\
    EMB33 & 587735664773431424 & 226.15445 & 48.73879 & 0.036099 & 10.53 & 9.45 ± 0.04 & 5 & —\\
    EMB34 & 587736584980463705 & 251.12815 & 19.94076 & 0.023001 & 10.29 & 9.36 & 3 & —\\
    EMB35 & 587736619863965778 & 245.33141 & 24.66665 & 0.037762 & 10.35 & — & 0 & —\\
    EMB36 & 587736541491233170 & 237.71840 & 5.32690 & 0.026031 & 9.66 & 9.57 ± 0.04 & 5 & —\\
    EMB37 & 587736808838594663 & 207.99749 & 13.96750 & 0.036694 & 11.27 & 9.99 ± 0.05 & 1 & —\\
    EMB38 & 587738410325573774 & 150.69248 & 12.27086 & 0.021737 & 9.72 & <8.59 & 5 & —\\
    EMB39 & 587738615415373880 & 170.94590 & 35.44231 & 0.034067 & 10.29 & 8.96 ± 0.05 & 5 & — \\
    EMB40 & 587738067810189520 & 133.03863 & 64.07869 & 0.036392 & 9.55 & <8.67 & 5 & —\\
    EMB41 & 587739130805420150 & 206.44644 & 34.49324 & 0.034790 & 9.54 & 9.33 ± 0.04 & 5 & —\\
    EMB42 & 587739303684931731 & 198.10781 & 34.12134 & 0.033809 & 9.89 & 8.85 ± 0.05 & 5 & —\\
    EMB43 & 587739504477077652 & 204.01718 & 30.14107 & 0.025928 & 9.64 & 9.07 ± 0.04$^\dagger$ & 5 & {J1336+3008$^2$}\\
    EMB44 & 587739166775574594 & 245.25338 & 21.16835 & 0.031025 & 10.38 & 8.61 ± 0.08 & 5 & —\\
    EMB45 & 587739609698140217 & 176.73898 & 32.65397 & 0.032757 & 9.61 & 8.24 ± 0.09 & 5 & —\\
    EMB46 & 587739646208770144 & 168.90291 & 30.42282 & 0.027881 & 9.65 & 9.58 ± 0.06 & 1 & —\\
    EMB47 & 587739646742429736 & 160.51722 & 29.55650 & 0.039814 & 10.06 & — & 0 & —\\
    EMB48 & 587739828743962776 & 228.00939 & 21.29817 & 0.015783 & 10.69 & 10.11 ± 0.17 & 1 & —\\
    EMB49 & 587739844856971464 & 241.64677 & 14.36721 & 0.032489 & 9.68 & 8.80 ± 0.05 & 5 & —\\
    EMB50 & 587739845393186912 & 240.21455 & 15.15126 & 0.033959 & 10.47 & 10.01 ± 0.06 & 1 & —\\
    EMB51 & 587741532777152653 & 149.75037 & 25.10314 & 0.022085 & 9.57 & 9.15 ± 0.07 & 1 & —\\
    EMB52 & 587741722823950491 & 195.50058 & 27.78273 & 0.023876 & 9.82 & <8.52 & 5 & —\\
    EMB53 & 587741815710744668 & 155.10063 & 21.35628 & 0.019190 & 10.02 & 8.76 ± 0.05 & 5 & —\\
    EMB54 & 587741721214386184 & 197.92175 & 26.39013 & 0.038107 & 9.98 & 9.91 ± 0.06 & 1 & —\\
    EMB55 & 587742010042744920 & 121.19475 & 10.77826 & 0.035262 & 10.87 & 10.18 ± 0.05 & 1 & —\\
    EMB56 & 587742189908983921 & 194.54164 & 24.34891 & 0.022669 & 9.58 & 9.38 ± 0.06 & 1 & — \\
    EMB57 & 587742576459251895 & 225.34016 & 15.24990 & 0.035373 & 9.91 & 9.17 ± 0.05 & 5 & —\\
    EMB58 & 587742589333340203 & 245.50907 & 9.88863 & 0.032715 & 10.48 & 8.63 ± 0.05 & 5 & —\\

    \hline
    \end{tabular}

   {\raggedright \textit{Note.} --- Columns are:
    1 -- EMBERS Survey ID;
    2 -- SDSS DR7 ObjID;
    3 -- Right ascension;
    4 -- Declination;
    5 -- Redshift from SDSS DR7 optical spectrum;
    6 -- Log of the stellar mass from the MPA/JHU catalog;
    7 -- Log of H$\text{\sc{i}}$ gas mass from \citetalias{ellisonLowRedshiftPoststarburst2025a};
    8 -- Source for H$\text{\sc{i}}$ data with {0=No available measurement}, 1=ALFALFA, 2=xGASS, 3=HI-MaNGA,
    4=FASHI, 5=FAST (New observations in \citetalias{ellisonLowRedshiftPoststarburst2025a}). HI-MaNGA and FASHI entries do not have associated errors; {9 -- Cross-referenced alternative ID of previous surveys from $^1$\citet{frenchDISCOVERYLARGEMOLECULAR2015g}, $^2$\citet{alataloSHOCKEDPOSTSTARBURSTGALAXY2016a}, and $^3$xCOLD GASS.}
    \raggedright
    $^\dagger$Mild RFI in spectrum; OK to use. 
    $^\ddagger$Strong RFI or Companion; No Measurement.
    }

    \label{tab:knowndata}
\end{table*}
\begin{table*}
    \addtocounter{table}{-1}
    \caption{Complete EMBERS sample galactic properties, H$\text{\sc{i}}$ information, and identifiers}
    \centering
    \setlength{\tabcolsep}{10pt}
    \begin{tabular}{ccccccccc}
    \hline 
    EMBERS ID & SDSS DR7 objID & RA & DEC  & $z$ &$\log \rm M_*$ & $\log \rm M_{\rm H\text{\sc{i}}}$ & H$\text{\sc{i}}$ Source & {Alt ID}\\
     &       & ($^{\circ}$) & ($^{\circ}$) &  & ($\rm M_{\sun}$) & ($\rm M_{\sun}$) & \\
    \hline
    EMB59 & 587742188827443225 & 174.95152 & 23.53247 & 0.030655 & 10.49 & <8.56 & 5 & —\\
    EMB60 & 587742189363527850 & 173.01533 & 23.70381 & 0.032286 & 10.19 & <8.65 & 5 & —\\
    EMB61 & 587742551762796902 & 238.81054 & 12.91632 & 0.032911 & 9.58 & 9.97 ± 0.05 & 1 & —\\
    EMB62 & 587742567860469784 & 159.24764 & 18.13767 & 0.020879 & 10.19 & 0.00$^\ddagger$ & 1 & —\\
    EMB63 & 587742616170266870 & 235.75415 & 16.98744 & 0.031407 & 9.87 & 8.95 ± 0.05 & 5 & —\\
    EMB64 & 587742627998531885 & 246.27553 & 8.38197 & 0.035113 & 10.29 & 9.71 ± 0.06 & 1 & —\\
    EMB65 & 587742628523475193 & 219.30721 & 14.66514 & 0.037896 & 10.88 & 9.62 ± 0.07 & 1 & —\\
    EMB66 & 587742644626260055 & 240.09445 & 12.75777 & 0.034376 & 10.12 & <8.57 & 5 & —\\
    EMB67 & 587745244159082657 & 136.46640 & 13.71746 & 0.027348 & 9.68 & 8.89 ± 0.05 & 5 & —\\
    EMB68 & 587745403073855572 & 137.87957 & 12.14780 & 0.029704 & 10.14 & <8.53 & 5 & —\\
    EMB69 & 588013382740279420 & 170.72341 & 51.34171 & 0.034165 & 9.85 & <8.69 & 5 & —\\
    EMB70 & 588015508189872263 & 346.93215 & -0.83848 & 0.032532 & 9.52 & 9.88 ± 0.05 & 5 & —\\
    EMB71 & 588015508208156794 & 28.63643 & -0.77010 & 0.016178 & 9.66 & <8.53 & 5 & —\\
    EMB72 & 588016891172618376 & 147.93714 & 35.62214 & 0.027049 & 10.56 & 9.89 ± 0.05 & 1 & {22822$^3$}\\
    EMB73 & 588017566019289130 & 167.82482 & 11.55440 & 0.038106 & 10.65 & — & 0 & {EAH07$^1$}\\
    EMB74 & 588016839633666164 & 123.32063 & 22.64830 & 0.022299 & 9.90 & 8.52 ± 0.06 & 5 & —\\
    EMB75 & 588017627780087861 & 227.22954 & 37.55827 & 0.029058 & 10.21 & 9.42 ± 0.05 & 5 & {EAH09$^1$}\\
    EMB76 & 588017724947759115 & 208.86976 & 6.59645 & 0.024082 & 9.83 & 9.58 ± 0.05 & 1 & —\\
    EMB77 & 588017979953643664 & 147.55031 & 34.69867 & 0.038380 & 9.95 & — & 0 & —\\
    EMB78 & 588018253751582792 & 223.73100 & 45.52406 & 0.036615 & 10.27 & — & 0 & —\\
    EMB79 & 588298661964546070 & 189.69499 & 46.46552 & 0.039636 & 9.82 & — & 0 & —\\
    EMB80 & 588848898848784582 & 202.57955 & -0.87069 & 0.037595 & 9.73 & 9.50 & 3 & —\\
    EMB81 & 588017992299380883 & 202.20521 & 10.38168 & 0.023096 & 9.59 & 9.64 ± 0.05 & 1 & —\\
    EMB82 & 588018055116095683 & 223.85983 & 44.89284 & 0.036599 & 9.64 & — & 0 & —\\
    EMB83 & 588018055117275233 & 226.98990 & 43.54559 & 0.039653 & 9.88 & — & 0 & —\\
    EMB84 & 588023046941245549 & 139.11457 & 19.92048 & 0.025981 & 9.99 & 9.58 ± 0.05 & 1 & —\\
    EMB85 & 588298664112881826 & 192.66358 & 47.93427 & 0.029111 & 9.63 & 9.34 ± 0.04 & 5 & —\\
    \hline
    \end{tabular}

   {\raggedright \textit{Note.} --- Columns are:
    1 -- EMBERS Survey ID;
    2 -- SDSS DR7 ObjID;
    3 -- Right ascension;
    4 -- Declination;
    5 -- Redshift from SDSS DR7 optical spectrum;
    6 -- Log of the stellar mass from the MPA/JHU catalog;
    7 -- Log of H$\text{\sc{i}}$ gas mass from \citetalias{ellisonLowRedshiftPoststarburst2025a};
    8 -- Source for H$\text{\sc{i}}$ data with {0=No available measurement}, 1=ALFALFA, 2=xGASS, 3=HI-MaNGA,
    4=FASHI, 5=FAST (New observations in \citetalias{ellisonLowRedshiftPoststarburst2025a}). HI-MaNGA and FASHI entries do not have associated errors; {9 -- Cross-referenced alternative ID of previous surveys from $^1$\citet{frenchDISCOVERYLARGEMOLECULAR2015g}, $^2$\citet{alataloSHOCKEDPOSTSTARBURSTGALAXY2016a}, $^3$xCOLD GASS.}
    \raggedright
    $^\dagger$Mild RFI in spectrum; OK to use. 
    $^\ddagger$Strong RFI or Companion; No Measurement.
    }

    \label{tab:knowndata2}
\end{table*}

We were granted time on the IRAM 30m telescope for CO observations of our PSBs over the course of four proposals (ID: 144-23, 077-24, 077-25, 180-24; PI: Jiménez-Donaire) for a total of {188.9 hours} on source including calibration overheads. The observations were undertaken from March 2024 to August 2025, with varying weather and telescope conditions. Individual observations varied from 18 minutes on source to 17 hours. Due to the faintness of the higher redshift, low stellar mass galaxies in the sample, we impose stricter mass and redshift cuts than \citetalias{ellisonLowRedshiftPoststarburst2025a} of log$_{10}$ M$_\ast$/M$_{\odot}$$>9.56$ and $0.01<z<0.0362$. Thus, of the 61 galaxies requiring new observations, we obtain measurements for {52} galaxies within our adjusted cuts, removing the 7 that are outside them. The remaining two galaxies unaccounted for correspond to EMB70, which was observed but lies outside of the stellar mass cut and does not meet our depth threshold, and EMB13, which was not observed due to time constraints. The relevant results of the observing campaign can be found in Table \ref{tab:psb_co10}. The mass and redshifts of all 86 galaxies in the EMBERS survey parent sample can be seen in Figure \ref{fig:mass redshift}. Galaxies with both atomic and molecular global data are shown as blue circles. Galaxies with only one gas phase measurement are shown as diamonds and those with no data are shown as light grey crosses. 

The EMBERS observing program was designed to match the sensitivity limits of the xCOLD GASS survey \citep{saintongeXCOLDGASSComplete2017a} such that we are sensitive to a molecular gas fraction of $f_{\rm{gas}}=1.5\%$ when log$_{10}$ M$_\ast$/M$_{\odot}$ $>10.5$ and $f_{\rm{gas}}=2.5\%$ at lower stellar masses. We use this detection threshold combined with the total stellar masses from the MPA/JHU catalog (see \citeauthor{kauffmann_stellar_2003} \citeyear{kauffmann_stellar_2003}) to estimate a minimum CO flux for each galaxy and thus a required integration time. We adopted the successful strategy employed by xCOLD GASS where the signal-to-noise (S/N) of each emission line was constantly monitored so that when securely detected CO(1--0) emission was achieved, integration stopped. 
\medskip 

All observations were undertaken with the Eight MIxer Receiver (EMIR; \citeauthor{carter_emir_2012} \citeyear{carter_emir_2012}) with the Fourier Transform Spectrometer (FTS) as the backend setup. The observations were performed in tracked, wobbler-switching mode with a 90 arcsecond throw between the ON and OFF positions. Given our redshift range, we were able to perform a single tuning of the combined EMIR E090+E230 receivers, permitting dual polarization observations simultaneously for both CO(1--0) and CO(2--1). The UO E090 sub-band was tuned to 114.36 GHz while the UO E230 sub-band was tuned to 226.52 GHz, with 8 GHz of bandwidth providing coverage of all lines in our survey. We leave the analysis of the CO(2--1) line for a future paper (Rasmussen et al., in prep). 

At the beginning of each observing run, pointing and focus corrections were applied using observations of planets or bright quasars. Focus corrections were performed roughly every three hours and at sunrise and sunset. The pointing of the telescope was corrected every $\sim $1--1.5 hours; when the weather was poor, pointing corrections were done more frequently. Weather conditions throughout the observing campaign differed significantly, with telescope system temperatures at 114.36 GHz ranging from $109  ~\rm K < ~T_{sys} ~< 405 ~K$. 

Each spectrum was reduced using the \texttt{CLASS} software \citep{pety_class_2005, CLASS}. The spectrum was centred on the expected position of the emission line using the galaxy's SDSS optical redshift and a $\pm$2000 kms$^{-1}$ region was extracted. The spectra were binned to a velocity resolution of 30 kms$^{-1}$. Individual scans for each galaxy were baseline-subtracted (zeroth order excluding a region matched to either the observed line width or, in the cases of a non-detection, the width of the H$\text{\sc{i}}$ line from \citetalias{ellisonLowRedshiftPoststarburst2025a}). Single-channel anomalous features outside of the line window were replaced with the background noise of the spectrum. All scans for a given galaxy were then combined into a final averaged spectrum where each scan was weighted by the inverse RMS noise outside of the central region. A final baseline subtraction was applied to the averaged spectrum (typically first-order but in the cases of distorted baselines, up to a third order was used). In cases where the edge of the two sidebands, upper-inner and upper-outer, fell within the extracted spectrum and the two sidebands were platformed with respect to each other, we fit the initial zeroth-order baselines to each sideband separately. 

\begin{figure*}
    \centering
    \newcommand{\vertspace}{-0.1cm} 

    \includegraphics[width=0.49\textwidth]{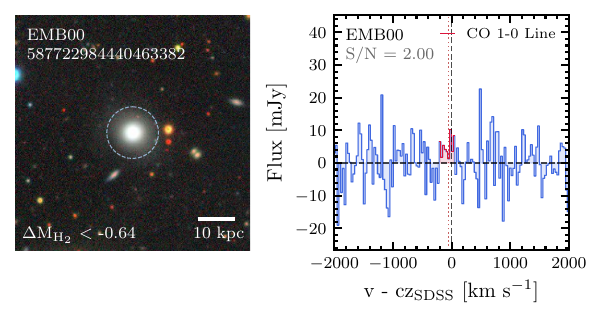}
    \includegraphics[width=0.49\textwidth]{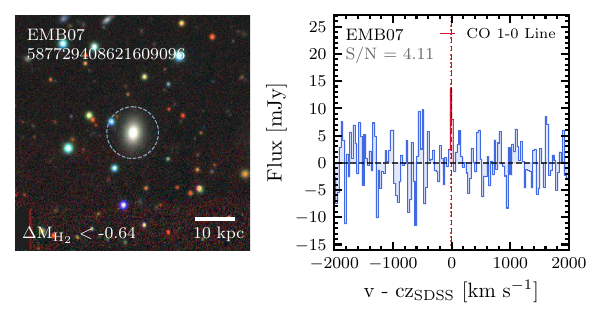}\\ \vspace{15pt}
    \includegraphics[width=0.49\textwidth]{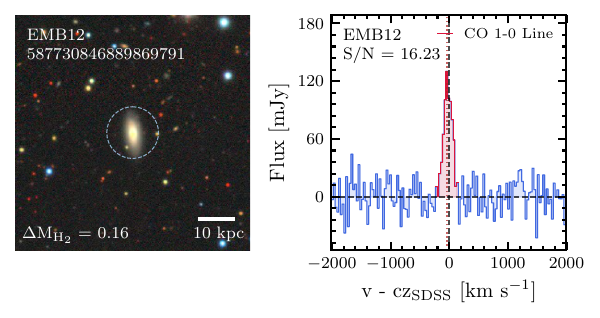}
    \includegraphics[width=0.49\textwidth]{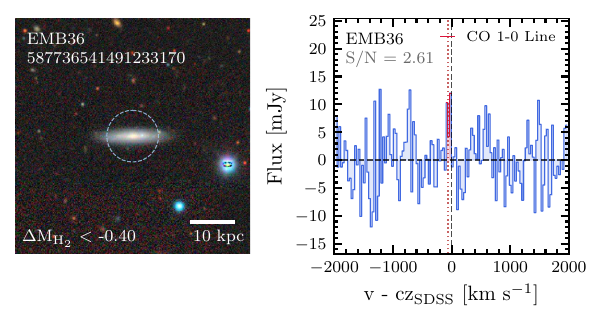}\\ \vspace{15pt}
    \includegraphics[width=0.49\textwidth]{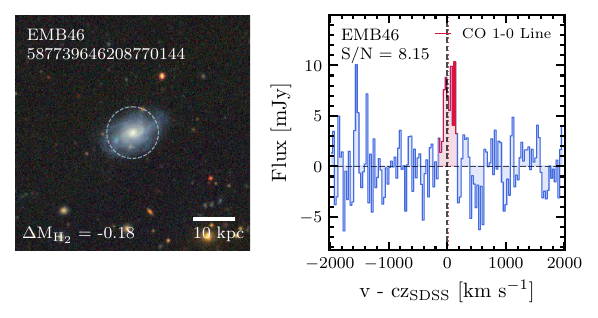}
    \includegraphics[width=0.49\textwidth]{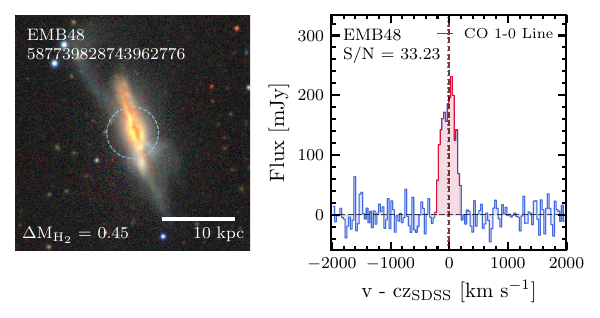}\\ \vspace{15pt}
    \includegraphics[width=0.49\textwidth]{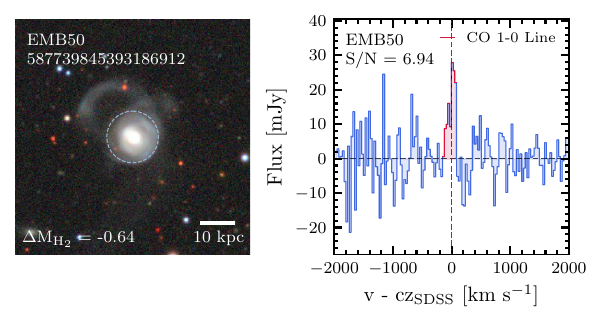}
    \includegraphics[width=0.49\textwidth]{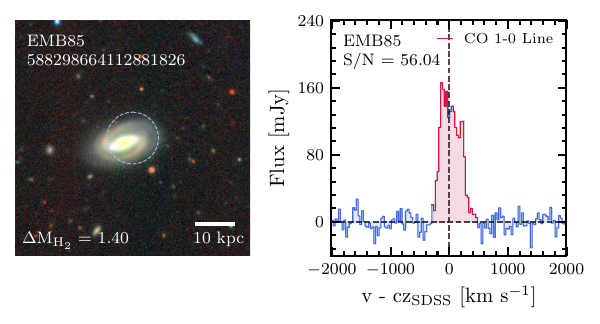}
    \caption{CO(1--0) spectra for eight galaxies chosen to demonstrate the diversity of PSB morphologies and molecular gas observations. \textit{First and third columns:} 100 arcsecond $\times$ 100 arcsecond DECaLS images with the 22 arcsecond beam size overlaid at 115 GHz for the IRAM 30m telescope \citep{deyOverviewDESILegacy2019}. The physical extent of 10 kpc is shown in the bottom right corner while the offset in molecular gas mass from SF galaxies in xCOLD GASS is shown in the bottom left corner. \textit{Second and fourth columns:} baseline subtracted CO(1--0) spectra with velocity resolution of 30 km/s centred on the SDSS optical redshift for each galaxy on its left. The region that is integrated over is highlighted in red (see Section \ref{sec:methods} for details). The vertical black dashed line denotes the expected velocity of the line from the redshift while the vertical red dotted line shows the velocity where half of the line flux lies on either side. A galaxy is considered securely detected if S/N$~\geq 5$. }
    \label{fig:psb_spectra_1}
\end{figure*}

Figure \ref{fig:psb_spectra_1} shows eight example CO(1--0) spectra from the sample taken with the IRAM 30m with matching identifiers to those in Table \ref{tab:psb_co10}. The galaxies in Figure \ref{fig:psb_spectra_1} were chosen to highlight the diversity of morphologies present in our sample of PSBs. EMB48 and EMB85 are the two most significantly detected galaxies in the sample and are outliers in many of the following figures. The remaining spectra can be found in Appendix \ref{app:additional spectra}. 

We compute the flux of a given line by identifying the peak channel within the central window excluded from the baseline subtraction and walk channel-by-channel outwards until a zero-point crossing on both sides of the line. The total flux, $\rm S_{\rm{CO}}$\footnote{Conversion factor from antenna temperature to flux density, $S/T^*_{\mathrm{a}}$, computed by linear interpolation of efficiencies at adjacent frequencies from here: \url{https://publicwiki.iram.es/Iram30mEfficiencies}. For CO(1--0) we use $6.1\ \mathrm{Jy}\ \mathrm{K^{-1}}$.} [Jy km s$^{-1}$], is then found by integrating the identified channels:

\begin{equation}
    \rm S_{\rm{CO}}=\int   \rm F_{\it{v}} ~d\it v,
\end{equation}
where $\rm F_\nu$ [mK] defines the region of the spectrum where the line is identified. Each spectrum was visually inspected to ensure that this method captures the emission line properly. We then compute the noise of each spectrum, $\sigma_{\rm CO}$ [mK], as the root mean square of the region that lies outside of the identified line. The width of the line, $\rm W_{85}$ [km s$^{-1}$], is computed as the region that contains 85 per cent of the total integrated line flux. We estimate the formal measurement error for observed line flux as done in \citet{saintongeCOLDGASSIRAM2011b} with:
\begin{equation}
    \varepsilon_{\rm obs} = \frac{\sigma_{\rm CO}W_{85}}{\rm \sqrt{W_{85}\Delta w_{ch}}},
\end{equation}
where $\Delta \rm w_{\rm ch}$ is the aforementioned $30\ \rm kms^{-1}$ channel width. Other contributions to the error budget include an 8 per cent flux calibration error for measurements at our frequencies with the IRAM 30m as is estimated by \citet{saintongeXCOLDGASSComplete2017a}. We do not include uncertainties on calibration.

We compute the signal-to-noise (S/N) ratio by the method of xCOLD GASS where:

\begin{equation}
    \rm{S/N} = \frac{S_{\rm{CO}}}{\varepsilon_{CO}}~.
    \label{eq:signal-to-noise}
\end{equation}
 We define a securely detected galaxy when S/N $\geq 5$ and record the integrated flux.  {We match the detection criteria of xCOLD GASS at the $5\sigma$ level such that a galaxy is securely detected in CO. For galaxies with $3<\rm S/N <5$, the strength of the emission lines are (visually) marginal, and as such, we impose a conservative detection definition at the $5\sigma$ level}. For galaxies where we do not achieve a secure detection in CO(1--0), we compute $3\sigma$ upper limits on the flux of the source following the technique in the xCOLD GASS survey described in \citet{saintongeXCOLDGASSComplete2017a} with the following equation:

\begin{equation}
    \rm S_{CO} = \int   \rm F_{\it{v}} d\it v = \rm 3 \cdot \sigma_{\rm CO} \cdot \sqrt{W_{CO} \cdot \Delta w_{ch}}~.
\label{eq:upper limit}
\end{equation}
$\rm S_{CO}$ [Jy km s$^{-1}$] is the $3 \sigma$ upper limit on flux from the source given noise $\sigma_{\rm CO}$ for a spectrum with channel width $\Delta \rm  w_{ch}$. $\rm W_{CO}$ is the assumed width of the undetected CO line which we take to be the H$\text{\sc{i}}$ linewidth from \citetalias{ellisonLowRedshiftPoststarburst2025a} if a given galaxy is detected in H$\text{\sc{i}}$ at the $5\sigma$ level. In the cases of an H$\text{\sc{i}}$ non-detection or non-observation, we match the procedure of xCOLD GASS and take $\rm W_{CO}$ to be 200 kms$^{-1}$ for log$_{10}$ M$_{\ast}$/M$_{\odot}$ < 10 and 300 kms$^{-1}$ for those with higher stellar mass.

CO luminosities of each source are computed in terms of the total integrated line flux using the method of \citet{solomon_LCO_1997} where:
\begin{equation}
    \rm L'_{CO} = 3.25 \times10^7 \ S_{CO} \ \nu_{obs}^{-2} \ D^2_L(\it z\rm)  \ (1+\it z \rm)^{-3}~.
\label{eq:luminosity}
\end{equation}
$ \mathrm{S}_{\mathrm{CO}} $ in units of Jy\,km\,s$^{-1}$ is the integrated line flux;\footnotemark[4]
$\nu_{\mathrm{obs}}$ is the observed frequency of the given line in GHz;
$\rm D_L$ is the luminosity distance in pc for our chosen cosmology. L$'_{\mathrm{CO}}$ is in units of  K km s$^{-1}$ pc$^2$ and is computed for CO(1--0) using Equation \ref{eq:luminosity} given either the integrated flux of a detection or the upper limit flux for non-detections from Equation \ref{eq:upper limit}. 

Converting the CO(1--0) luminosity of a PSB to the inferred total molecular gas mass (hereby denoted M$_{\rm H_2}$) is done by:
\begin{equation}
    \rm \ M_{\rm H_2} = \alpha_{CO} \cdot L'_{CO(1-0)} ~.
    \label{eq:alpha L}
\end{equation}
$\alpha_{\rm CO}$ is the assumed conversion between CO gas mass and total molecular gas mass with units of $\rm M_\odot(\mathrm{K\ km\ s^{-1}\ pc^2})^{-1}$  {which we apply to both new and archival measurements}. In recent years, conversion factor parametrizations that depend on SFR and metallicity have been proposed \citep{sandstrom_2013,  accurso_alpha_2017,chiang_2024,schinnerer_leroy_2024rev}. As the majority of star formation rate tracers for PSBs are typically unreliable and many of our PSBs lack emission lines to derive gas-phase metallicities, more complex prescriptions of the conversion factor, $\alpha_{\rm CO}$, like those in \citet{accurso_alpha_2017} are unavailable  {directly} for our sample. As such, we adopt a constant value of the conversion factor of $\alpha_{\rm CO} = 4.3 \rm \ M_{H_2}/L_{CO}'$ consistent with Milky-Way molecular clouds and the Local Group \citep{bolatto_alpha_2013}. However, in Section \ref{sec:alpha} we explore how plausible changes to $\alpha_{\rm CO}$ might affect our results, finding that our conclusions are qualitatively robust to these uncertainties.

Given the redshifts of our PSBs (in Table \ref{tab:knowndata}) and the beam size at 114.32 GHz for the IRAM 30m telescope of $\sim 22$ arcseconds, our observations capture CO(1--0) emission within physical diameters between 7.1 kpc and 15.4 kpc with a median size of 13.5 kpc. Recent resolved studies of molecular gas in PSBs have shown that the CO emission is highly compact. \citet{smercinaFallResolvingMolecular2022} find that the CO half-mass radius, $\rm R_{50}$, for their PSBs ranges from 0.11 kpc to 0.72 kpc with a median of 0.22 kpc, while \citet{otterResolvedMolecularGas2022} similarly find that $\rm R_{50}$ for their PSBs ranges from 0.6 kpc to 3.9 kpc with a median of 1.2 kpc, with one of the sources having unresolved CO emission at $2.5$ arcsecond resolution. The median R$_{50}$ for all resolved PSBs in the literature is $\sim 0.7$ kpc. We thus argue that it is likely that the beam at 114.32 GHz captures the entirety of CO emission in our PSBs for CO(1--0). As such, any attempted aperture corrections would cause us to artificially overestimate the CO flux for each galaxy. Additionally, the majority of the PSBs' total angular sizes are smaller than the 22 arcsecond beam as can be seen in Figures \ref{fig:psb_spectra_1} and Appendix \ref{app:additional spectra}. We note that in a handful of low redshift cases (EMB09, EMB48, EMB55, and EMB62 specifically), the beam size is not sufficient to cover the spatial extent of the galaxy. We discuss the implications for galaxies larger than the beam in Section \ref{sec:deltas}.


With the completion of observations for our PSB sample, we have obtained new global measurements of CO(1--0) with the IRAM 30m for {52} galaxies. Only one galaxy, EMB13, within our mass and redshift cuts defined in Section \ref{sec:sample} went unobserved  {or lacked archival CO observations. We find one PSB CO measurement for EMB72 in xCOLD GASS and EMB03 and EMB43 from the \citet{alataloSHOCKEDPOSTSTARBURSTGALAXY2016a} SPOGS sample. Additionally, EMB11, EMB19, EMB22, EMB32, EMB73, and EMB74 CO measurements are taken from the \citet{frenchDISCOVERYLARGEMOLECULAR2015g} sample. The cross-referenced identifiers for the 9 archival measurements can be found in Table \ref{tab:knowndata}. Of these 9 archival galaxies, 7 are securely detected in CO at the $5\sigma$ level, with upper limits on the two non-detections from \citet{frenchDISCOVERYLARGEMOLECULAR2015g} of EMB22 and EMB73 computed in the same fashion as Equation \ref{eq:upper limit}. With the 9 additional archival measurements, the total CO sample is 61 PSBs. Of these {61} measurements, {58} have reliable H$\text{\sc{i}}$ observations from \citetalias{ellisonLowRedshiftPoststarburst2025a} allowing us to do a complete and systematic census of both the atomic and molecular gas for a large and representative population of PSBs. }


\begin{table*}
    \centering
    \setlength{\tabcolsep}{12pt}
    \footnotesize
    \caption{CO(1--0) observations for the EMBERS sample}
    \begin{tabular}{ccccccccc}
    \hline
    ID & $ \int   \rm F_{\it{v}} d\it v$ & RMS & S/N & $\rm W_{85}$ & $\rm L'_{\rm CO}$ & $\log \rm M_{\rm H_2}$ & $\rm T_{\rm on}$ & CO Source \\
       & (K km s$^{-1}$) & (mK) &  & (km s$^{-1}$) & ($10^7$ K km s$^{-1}$ pc$^2$) & ($\rm M_\odot$) & (hrs) & \\
    \hline
    EMB00 & 0.22 ± 0.11 & 1.20 & 2.00 & 280 & $<$8.34 & $<$8.55 & 7.20 &4\\
    EMB02 & 0.30 ± 0.05 & 0.74 & 6.20 & 147 & 7.39 ± 1.33 & 8.50 ± 0.18 & 5.20 &4\\
    EMB03 & — & — & — & — & — & 8.99 ± 0.20 & — & 2 \\
    EMB07 & 0.12 ± 0.03 & 0.71 & 4.11 & 55 & $<$3.37 & $<$8.16 & 5.75 &4\\
    EMB08 & 0.06 ± 0.08 & 1.20 & 0.91 & 119 & $<$10.96 & $<$8.67 & 2.10 &4\\
    EMB09 & 4.28 ± 0.40 & 2.63 & 21.39 & 193 & 151.61 ± 14.05 & 9.81 ± 0.09 & 0.60 &4\\
    EMB10 & 0.40 ± 0.10 & 1.26 & 4.29 & 182 & $<$6.15 & $<$8.42 & 2.20 &4\\
    EMB11 & — & — & — & — & — & 8.60 ± 0.25 & — & 1 \\
    EMB12 & 3.60 ± 0.36 & 2.81 & 16.23 & 208 & 103.37 ± 10.44 & 9.65 ± 0.10 & 0.45 &4\\
    EMB14 & 0.51 ± 0.11 & 1.30 & 5.12 & 195 & 14.95 ± 3.15 & 8.81 ± 0.21 & 3.00 &4\\
    EMB18 & 0.06 ± 0.04 & 1.19 & 1.68 & 30 & $<$4.35 & $<$8.27 & 4.80 &4\\
    EMB19 & — & — & — & — & — & 8.74 ± 0.33 & — & 1 \\
    EMB20 & 0.95 ± 0.14 & 1.43 & 7.81 & 240 & 16.01 ± 2.42 & 8.84 ± 0.15 & 1.95 &4\\
    EMB21 & 0.08 ± 0.03 & 0.47 & 3.26 & 95 & $<$1.60 & $<$7.84 & 6.35 &4\\
    EMB22 & — & — & — & — & — & $<$8.67 & — & 1 \\
    EMB23 & 0.10 ± 0.12 & 1.33 & 0.87 & 259 & $<$4.02 & $<$8.24 & 1.45 &4\\
    EMB24 & 1.00 ± 0.16 & 1.59 & 7.54 & 233 & 27.80 ± 4.31 & 9.08 ± 0.15 & 1.37 &4\\
    EMB25 & 0.27 ± 0.05 & 0.61 & 5.42 & 225 & 7.06 ± 1.42 & 8.48 ± 0.20 & 5.80 &4\\
    EMB26 & 0.04 ± 0.03 & 0.44 & 1.38 & 130 & $<$4.02 & $<$8.24 & 10.10 &4\\
    EMB28 & 0.05 ± 0.02 & 0.49 & 2.27 & 63 & $<$2.73 & $<$8.07 & 7.20 &4\\
    EMB32 & — & — & — & — & — & 9.15 ± 0.16 & — & 1 \\
    EMB33 & 0.94 ± 0.12 & 0.98 & 9.43 & 343 & 34.01 ± 4.52 & 9.17 ± 0.13 & 3.10 &4\\
    EMB34 & 2.04 ± 0.23 & 1.89 & 12.61 & 244 & 29.85 ± 3.36 & 9.11 ± 0.11 & 1.20 &4\\
    EMB36 & 0.13 ± 0.05 & 0.86 & 2.61 & 114 & $<$4.10 & $<$8.25 & 4.20 &4\\
    EMB38 & 0.04 ± 0.02 & 0.72 & 1.85 & 30 & $<$2.19 & $<$7.97 & 5.55 &4\\
    EMB39 & 0.21 ± 0.07 & 1.27 & 2.99 & 101 & $<$10.67 & $<$8.66 & 1.70 &4\\
    EMB42 & 0.66 ± 0.09 & 0.77 & 9.74 & 258 & 20.96 ± 2.73 & 8.95 ± 0.13 & 5.10 &4\\
    EMB43 & — & — & — & — & — & 8.72 ± 0.04 & — & 2 \\
    EMB44 & 0.35 ± 0.09 & 1.04 & 3.97 & 242 & $<$5.34 & $<$8.36 & 3.40 &4\\
    EMB45 & 0.04 ± 0.02 & 0.39 & 2.55 & 62 & $<$1.68 & $<$7.86 & 12.10 &4\\
    EMB46 & 0.31 ± 0.05 & 0.46 & 8.15 & 229 & 6.70 ± 0.98 & 8.46 ± 0.15 & 6.45 &4\\
    EMB48 & 9.86 ± 0.84 & 3.13 & 33.23 & 300 & 436.93 ± 37.35 & 10.27 ± 0.09 & 0.30 &4\\
    EMB49 & 0.02 ± 0.05 & 0.81 & 0.40 & 127 & $<$4.00 & $<$8.24 & 4.80 &4\\
    EMB50 & 0.59 ± 0.10 & 1.19 & 6.94 & 169 & 18.91 ± 3.12 & 8.91 ± 0.16 & 2.40 &4\\
    EMB51 & 0.09 ± 0.03 & 0.60 & 3.18 & 83 & $<$2.13 & $<$7.96 & 6.80 &4\\
    EMB52 & 0.04 ± 0.02 & 0.79 & 1.69 & 30 & $<$2.91 & $<$8.10 & 5.15 &4\\
    EMB53 & 0.76 ± 0.13 & 1.63 & 6.87 & 155 & 7.75 ± 1.29 & 8.52 ± 0.17 & 1.20 &4\\
    EMB55 & 2.17 ± 0.31 & 2.14 & 8.63 & 458 & 75.18 ± 10.59 & 9.51 ± 0.14 & 0.44 &4\\
    EMB56 & 0.98 ± 0.16 & 1.44 & 6.99 & 316 & 13.94 ± 2.28 & 8.78 ± 0.16 & 0.90 &4\\
    EMB57 & 0.39 ± 0.07 & 0.87 & 6.69 & 151 & 13.63 ± 2.31 & 8.77 ± 0.17 & 3.60 &4\\
    EMB58 & 0.47 ± 0.10 & 1.26 & 5.21 & 171 & 14.04 ± 2.92 & 8.78 ± 0.21 & 1.70 &4\\
    EMB59 & 0.33 ± 0.12 & 2.45 & 2.84 & 75 & $<$14.86 & $<$8.81 & 0.28 &4\\
    EMB60 & 0.60 ± 0.13 & 1.93 & 5.16 & 121 & 17.39 ± 3.65 & 8.87 ± 0.21 & 0.77 &4\\
    EMB61 & 0.07 ± 0.02 & 0.52 & 3.49 & 52 & $<$3.90 & $<$8.23 & 17.40 &4\\
    EMB62 & 3.70 ± 0.40 & 2.61 & 13.62 & 359 & 44.58 ± 4.84 & 9.28 ± 0.11 & 0.29 &4\\
    EMB63 & 0.38 ± 0.07 & 0.99 & 6.06 & 136 & 10.48 ± 1.92 & 8.65 ± 0.18 & 3.20 &4\\
    EMB64 & 0.38 ± 0.10 & 1.44 & 3.87 & 157 & $<$10.03 & $<$8.63 & 1.50 &4\\
    EMB66 & 0.29 ± 0.07 & 0.82 & 4.29 & 229 & $<$6.24 & $<$8.43 & 3.60 &4\\
    EMB67 & 0.44 ± 0.07 & 0.76 & 7.82 & 181 & 9.10 ± 1.37 & 8.59 ± 0.15 & 2.35 &4\\
    EMB68 & 0.11 ± 0.05 & 1.05 & 2.45 & 56 & $<$5.98 & $<$8.41 & 1.20 &4\\
    EMB69 & 1.05 ± 0.13 & 1.64 & 10.26 & 130 & 34.04 ± 4.29 & 9.17 ± 0.13 & 0.75 &4\\
    EMB70 & 0.024 ± 0.027 & 0.60 & 1.41 & 53 & $<$4.30 & $<$8.27 & 8.4 & 4 \\
    EMB71 & 0.14 ± 0.13 & 1.52 & 1.08 & 228 & $<$2.55 & $<$8.04 & 1.50 &4\\
    EMB72 & — & — & — & — & — & 9.39 ± 0.17 & — & 3 \\
    EMB73 & — & — & — & — & — & 8.65 ± 0.37 & — & 1 \\
    EMB74 & 0.16 ± 0.09 & 1.21 & 1.69 & 195 & $<$2.82 & $<$8.08 & 1.35 &4\\
    EMB75 & — & — & — & — & — & $<$9.18 & — & 1 \\
    EMB76 & 1.58 ± 0.18 & 1.74 & 12.02 & 192 & 25.47 ± 2.94 & 9.04 ± 0.12 & 0.75 &4\\
    EMB81 & 0.25 ± 0.05 & 0.58 & 5.69 & 187 & 3.65 ± 0.70 & 8.20 ± 0.19 & 6.30 &4\\
    EMB84 & 0.59 ± 0.09 & 0.89 & 7.71 & 243 & 11.00 ± 1.68 & 8.67 ± 0.15 & 3.05 &4\\
    EMB85 & 10.54 ± 0.86 & 1.66 & 56.04 & 426 & 248.33 ± 20.35 & 10.03 ± 0.08 & 0.61 &4\\
    \hline
    \end{tabular}

        {\raggedright \scriptsize \textit{Note.} --- Columns are:
    1 -- EMBERS sample identifier;
    2 -- Integrated CO(1-0) line flux;
    3 -- RMS of spectrum;
    4 -- S/N of spectrum. If less than 5, upper limits are computed.
   4-- linewidth that contains 85 \% of the flux. If W$_{85}$=30 km s$^{-1}$, only one channel is included;
    6 -- CO line luminosity;
    7 -- Log of molecular gas mass inferred from CO emission;
    8 -- IRAM 30m integration time;
    9 -- Source for CO data. 1=From \cite{frenchDISCOVERYLARGEMOLECULAR2015g}, 
    2=From \cite{alataloSHOCKEDPOSTSTARBURSTGALAXY2016a}, 
    3=From xCOLDGASS, 4=New observations with IRAM 30m. Cross-referenced IDs for archival measurements can be found in Section \ref{sec:observations}.}
    
    \label{tab:psb_co10}
\end{table*}

\clearpage
\section{results}
\label{sec:results}

With observations of CO(1--0) obtained in 61 EMBERS PSBs, we now present the results of their bulk molecular gas properties and compare them to a control sample of normal (non-PSB) galaxies.
\subsection{Bulk molecular gas reservoirs}
\label{sec:comparison}
Evaluating the molecular gas properties of PSBs with respect to normal galaxies requires a representative comparison sample. We utilize the xCOLD GASS survey \citep{saintongeXCOLDGASSComplete2017a} which measured the CO emission and inferred molecular gas masses for 532 galaxies in the redshift range of $0.01 < z < 0.05$ and mass range of log$_{10}$ M$_{\ast}$/M$_{\odot}$ $>9$ for both CO(1--0) and CO(2--1). The xCOLD GASS survey provides us with sufficient mass and redshift coverage to appropriately compare the results of our PSB galaxies with. As described in the previous section, the sensitivity of the EMBERS IRAM 30m survey was also designed to match that of xCOLD GASS. 

\begin{figure*}
  \centering
  \includegraphics[width=0.475\textwidth]{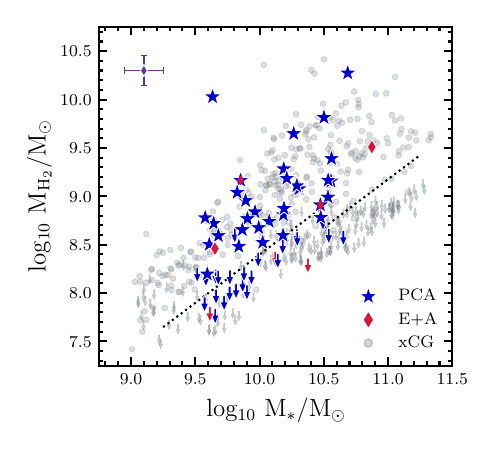}\hspace{-0.0\textwidth}
  \includegraphics[width=0.475\textwidth]{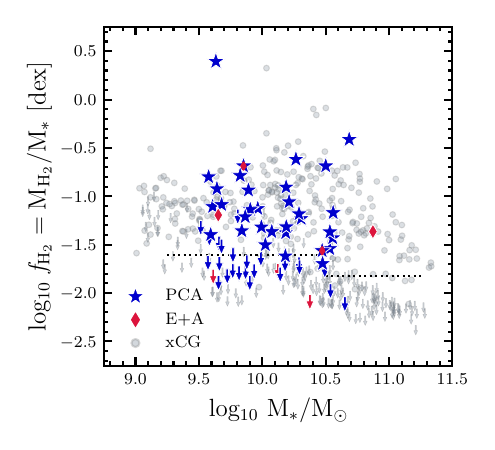}
  \caption{The total inferred molecular gas mass (left) and gas fraction (right) as a function of stellar mass for EMBERS PSBs coloured as blue stars for PCA selected galaxies and as red diamonds for E+A PSBs. In grey are the aperture corrected measurements from xCOLD GASS \citep{saintongeXCOLDGASSComplete2017a}. Upper limits in both samples are denoted as downward arrows. All data points in the figure assume a CO-to-H$_2$ ratio of $\alpha_{\mathrm{CO}} = 4.3\ \mathrm{M}_\odot\,(\mathrm{K\,km\,s^{-1}\,pc^{2}})^{-1}$. The solid black lines denote the survey sensitivity threshold corresponding to 1.5 per cent gas fraction for log$_{10}$ M$_{\ast}$/M$_{\odot}$ > 10.5  and 2.5 per cent for log$_{10}$ M$_{\ast}$/M$_{\odot}$ < 10.5. The typical uncertainties on stellar and molecular gas masses for the PSBs are shown in the top left of the left panel. Although PSBs are in the process of quenching, we detect molecular gas in the majority of our sample, with bulk gas reservoirs spanning $\sim 2.5$ dex in gas fraction ranging from 0.8 per cent to $\sim 250$ per cent.  }
  \label{fig:gass distributions}
\end{figure*}

As we have previously argued, the majority of our PSBs are smaller than the 22 arcsecond beam of the IRAM 30m at $\sim 115 \rm GHz$ and the CO emission is likely highly compact. Conversely, as shown by \citet{wylezalek_2022}, some extended CO(1--0) emission is lost at the 22 arcsecond scale for normal galaxies. Thus, the appropriate comparison sample for \textit{total} gas masses is then the aperture corrected xCOLD GASS. PSB non-detections that lie above our designed detection threshold are due to a combination of the prohibitive integration times required for the lower mass and higher redshift galaxies in the sample and the scheduled observing times. The presence of low-mass non-detections above a gas fraction of 2.5 per cent is consistent with the behaviour of the xCOLD GASS sample as well, and so we include these observations moving forward.

In Figure \ref{fig:gass distributions} we plot the inferred molecular gas mass from CO(1--0) emission in the left-hand panel as well as the gas fraction ($f_{\rm H_2}=\rm M_{H_2}/M_{\ast}$) in the right-hand panel, both as a function of stellar mass. The detection thresholds of xCOLD GASS are shown as dotted black lines in both panels of Figure \ref{fig:gass distributions}. We make the distinction between PCA selected PSBs (blue stars) as well as the traditional E+A selected PSBs (red diamonds) where galaxies selected by both methods are denoted with the overlay of the two markers. In galaxies that do not achieve a S/N  greater than five, as defined in Equation \ref{eq:signal-to-noise}, we compute upper limits with Equation \ref{eq:upper limit} and plot them as downward arrows. The median uncertainty in both gas mass and stellar mass is shown as a purple error bar in the top left corner of Figure \ref{fig:gass distributions}. The grey points in the background are aperture corrected measurements of the xCOLD GASS sample using the same conversion factor, $\alpha_{\rm CO} = 4.3 \rm \ M_{H_2}/L_{CO}'$, as we adopt for the PSBs.

As demonstrated by Figure \ref{fig:gass distributions}, the majority of PSBs still maintain ample gas reservoirs despite their rapidly declining star formation rates. We securely detect CO emission in 34 galaxies of the 61 with observations, with the fraction of detected galaxies increasing strongly with increasing stellar mass. We find that both the total gas mass and gas fractions in our PSBs are diverse, spanning over 2.5 dex from 0.8 per cent to over 200 per cent, and occupy largely the same parameter space as the xCOLD GASS representative galaxies. Although EMBERS is dominated by PCA selected galaxies (due to the more inclusive selection criteria), we find that PCA and E+A galaxies are distributed similarly among non-detections, detections, and gas-rich galaxies. 

We make two additional qualitative observations of the distributions of our PSBs. The first is that there is a significant population ($\sim 10$ galaxies) of our PSBs that are categorically gas-rich in comparison to the bulk of the xCOLD GASS distributions, with molecular gas fractions $>10$ per cent of their stellar mass. In future sections, this elevated gas fraction will be compared quantitatively to the xCOLD GASS population. Two galaxies in particular, EMB48 (log$_{10}$ M$_{\ast}$/M$_{\odot}$ $=10.69$) and EMB85 (log$_{10}$ M$_{\ast}$/M$_{\odot}$ $=9.63$) have very large molecular gas masses of log$_{10}$ M$_{\rm H_2}$/M$_{\odot}$ $>10$. At low stellar masses, we also note that there is a population of sufficiently deep non-detections consistent with a population of molecular gas \textit{depleted} PSBs. We apply a stacking routine to quantify this further in Section \ref{sec:stacking}.

The importance of a representative sample of PSBs with suitably deep observations is highlighted by the existence of the low M$_\ast$ population with deep CO non-detections. At higher stellar masses, our observations are consistent with the results of \citet{frenchDISCOVERYLARGEMOLECULAR2015g} and \citet{rowlands_psbs_2015} which both have previously discovered significant gas reservoirs in their populations of PSBs (albeit with different selection methods and stellar masses/redshifts). In both previous studies, the large majority of sources have stellar masses log$_{10}$ M$_{\rm \ast}$/M$_{\odot}$ $>10$ and are detected in CO(1--0) in 17/32 cases for \citet{frenchDISCOVERYLARGEMOLECULAR2015g} and 10/11 cases for \citet{rowlands_psbs_2015}. Both sets of observations are typically not sensitive to gas masses less than log$_{10}$ M$_{\rm H_2}$/M$_{\odot}$ $=8.5$, which is the regime where the majority of our non-detections lie. The homogeneous nature of our survey design, matched to xCOLD GASS, allows us to make a direct comparison to the molecular gas properties of a more representative set of galaxies and mitigates any sample selection bias.

In summary, we securely detect CO(1--0) in 34/61 PSBs. Additionally, Figure \ref{fig:gass distributions} demonstrates qualitatively that PSBs selected via both E+A and PCA methods have molecular gas masses and gas fractions that span the range of our comparison sample (xCOLD GASS). In the following sections, we now extend the analysis to quantitative comparisons in detection fraction and gas fraction between the two samples.

\begin{figure*}
  \centering
  \includegraphics[width=0.43\textwidth]{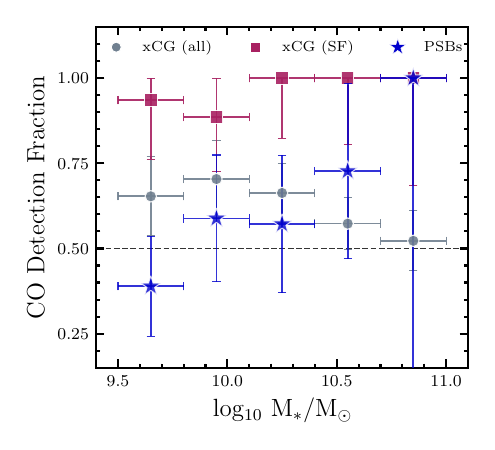}\hspace{0.025\textwidth}
  \includegraphics[width=0.43\textwidth]{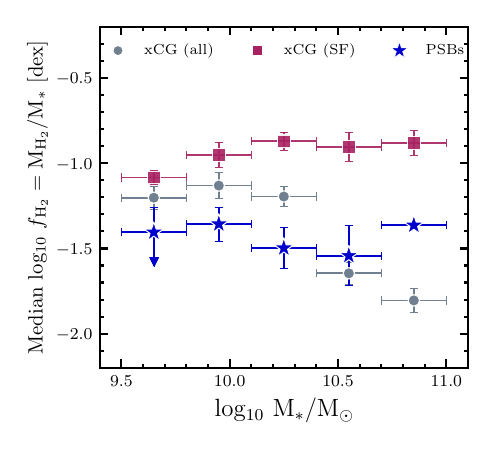}
  \caption{\textit{Left:} fraction of galaxies where CO is detected at the $\rm S/N>5$ level for our PSBs (blue), the entire xCOLD GASS sample (grey), and a BPT selected star-forming subset of xCOLD GASS (violet) in bins of stellar mass. The horizontal dashed line corresponds to a detection fraction of 50$\%$ where a robust median value can still be computed. Error bars are the binomial error for each bin. Low stellar mass PSBs are detected at a rate consistently lower than the xCOLD GASS comparison sample while higher mass PSBs are detected at a consistent or elevated rate. PSBs are detected at significantly lower levels than SF galaxies. \textit{Right:} median gas fraction in bins of stellar mass for the same three samples of galaxies as the left panel. The downwards arrow denotes an upper limit due to a detection fraction $<50\%$ for that bin. Errors on each bin are the standard error on the median. Similar to detection fractions, the median gas mass is lower than the comparison sample for low stellar mass PSBs. The PSBs do not experience the downturn shown by the xCOLD GASS comparison, caused by the increase in quenched population in normal galaxies. When the quenched population is removed from the comparison, as is done for the SF subset of xCOLD GASS, our PSBs are consistently between $\sim$0.3 -- 0.6 dex lower in gas fraction than the assumed star-forming progenitors.}
  \label{fig:fractions}
\end{figure*}
\subsection{Rate of detections and median gas fractions}
\label{sec:gas fractions}
\subsubsection{Detection fractions}
To directly compare the EMBERS PSB sample to the xCOLD GASS survey, we first look to quantify the rate of detections between the two samples. We apply the same detection threshold to xCOLD GASS as for the PSBs, i.e. $\rm S/N>5$. We combine the statistics from the PCA-selected galaxies and traditional E+A PSBs for the remainder of the analysis as the gas properties between the two samples do not appear to differ systematically. Prior to the shutdown of star formation seen in PSBs, it is likely the case that they were once normal, SF galaxies and, as such, just the SF galaxies in xCOLD GASS may act as a more suitable comparison sample. We select this SF subset of xCOLD GASS galaxies using the \citet{kauffmann_bpt_2003} cut in [OIII]/H$\beta$ -- [NII]/H$\alpha$ space of the BPT diagram (Baldwin, Phillips $\&$ Terlevich; \citeauthor{BPT_1981} \citeyear{BPT_1981}), requiring a signal-to-noise ratio greater than 3 for each emission line. 
The left panel of Figure \ref{fig:fractions} shows the detection fraction of our PSBs and the comparison sample of xCOLD GASS and the SF subset as a function of stellar mass, in bins of 0.3 dex. The horizontal black line corresponds to a detection fraction of 50 per cent; if a point lies above this line, we are able to compute a robust median value that is not an upper limit. Errors on each point correspond to the $1\sigma$ binomial error. Our PSBs are shown as blue stars, while the xCOLD GASS samples are shown as grey circles (all) and violet squares (SF). 

To orient ourselves, we note that when considering the xCOLD GASS sample in its entirety, CO detection fractions always lie above 50 per cent. As stellar mass increases, so does the increased fraction of quenched galaxies, and we see a downturn in detection fraction. When considering just SF xCOLD GASS galaxies, molecular gas is detected in almost all cases, with a slight decrease at low stellar masses likely due to low required sensitivities to achieve a detection. 

Figure \ref{fig:fractions} demonstrates a number of key conclusions of this work when we compare the detection fraction of our PSBs to xCOLD GASS. Firstly, the majority of our non-detections are at low stellar masses, resulting in a sub-50 per cent detection fraction in the lowest bin of $9.5<$log$_{10}~$M$_{\ast}$/M$_{\odot}<9.8$. Additionally, we consistently find lower CO detection fractions compared to the full xCOLD GASS sample until log$_{10}~$M$_{\ast}$/M$_{\odot}$ $>10.4$ where we find elevated rates of the detection of CO. Contrary to the statistics of the complete xCOLD GASS sample, we find that the CO detection fraction in PSBs increases with stellar mass, rather than decreases. When we compare the rates of CO detection to just SF galaxies, we find significantly lower detection fractions across stellar mass bins until the highest M$_{\ast}$ bin, where our statistics are small. Therefore, despite detecting CO in $>50$ per cent of EMBERS PSBs (34/61), when compared specifically to their presumed progenitor population of SF galaxies, the PSBs are detected at significantly lower rates.

\subsubsection{Median gas fractions}
As shown in the right panel of Figure \ref{fig:fractions}, in bins where the detection fraction is greater than half, we can compute median gas fractions for both PSBs and our comparison samples. The errors on each point correspond to the $1\sigma$ standard error on the median. For the one PSB bin where we do not detect CO in greater than half of the galaxies, we denote an upper limit with a downward arrow. 

The median gas fractions of the full xCOLD GASS sample remain relatively constant for low stellar masses, with a sharp downturn when log$_{10}~$M$_{\ast}$/M$_{\odot}$ $>10.4$. As with the detection fractions, this is related to the increase in the quenched fraction at higher stellar mass. The increase in quenched fraction can be seen explicitly as the difference between the median gas fractions between the full xCOLD GASS sample and just the SF subset widens considerably with increasing stellar mass. The median gas fraction of our PSBs remains relatively constant across 1.5 dex in stellar mass, with a molecular gas fraction of $\sim 3$ per cent. When we compare the median gas fractions of our PSBs to the full xCOLD GASS sample, we find a consistent decrement across the low mass regime of $\sim 0.2-0.3$ dex. The difference then plateaus at higher stellar masses as xCOLD GASS turns downward. Compared to their SF progenitors, the typical PSB is significantly and consistently between $\sim 0.3-0.6$ dex deficient in molecular gas, with the gap widening with an increase in stellar mass until the last, poorly populated bin. 

\medskip 

In summary of this sub-section, we find that the CO detection fraction and median molecular gas mass for PSBs are both consistently lower than their star-forming progenitors across a large stellar mass range. We can then say that, despite frequent detections of CO(1--0), PSBs, as a population, are nonetheless molecular gas deficient. However, Figure \ref{fig:gass distributions} also showed that the gas properties of individual PSBs are diverse.  {The results from Figures \ref{fig:gass distributions} and  \ref{fig:fractions} are comparable to the H$\text{\sc{i}}$ results from \citetalias{ellisonLowRedshiftPoststarburst2025a}, where the atomic gas in the population of EMBERS PSBs is relatively depleted compared to SF galaxies. However, individual PSBs show significant diversity. That being said, molecular gas detections are less common than atomic gas detections at lower stellar masses, while the gap, or depletion, in global gas content between SF galaxies and our PSBs is much larger at high stellar mass in the molecular phase as opposed to atomic. Access to both phases allows us to compare the ratio of the molecular and atomic gas phases in PSBs for the first time.  }


\subsection{Molecular-to-Atomic Gas Ratio}
\label{sec:gas ratios}

 {We have established in Section \ref{sec:results} thus far, as well as in \citetalias{ellisonLowRedshiftPoststarburst2025a}, that both the atomic and molecular gas contents of EMBERS PSBs are diverse. Both phases are necessary in a galaxy to form stars and sustain new stellar growth as the atomic gas reservoir cools to form molecular gas \citep{leroy_2008,larson_2016}. PSBs might therefore be expected to show signs of the conversion between the atomic and molecular gas reservoirs to break down if rapid-quenching is initiated by an inability to convert diffuse atomic hydrogen to denser, colder molecular gas. The ratio between the global atomic and molecular gas content can describe the efficiency of this atomic-to-molecular gas conversion \citep{niankun_atomic-molecular_2024}. We define this metric as: }
\begin{figure}
\centering
    \includegraphics[width=0.49\textwidth]{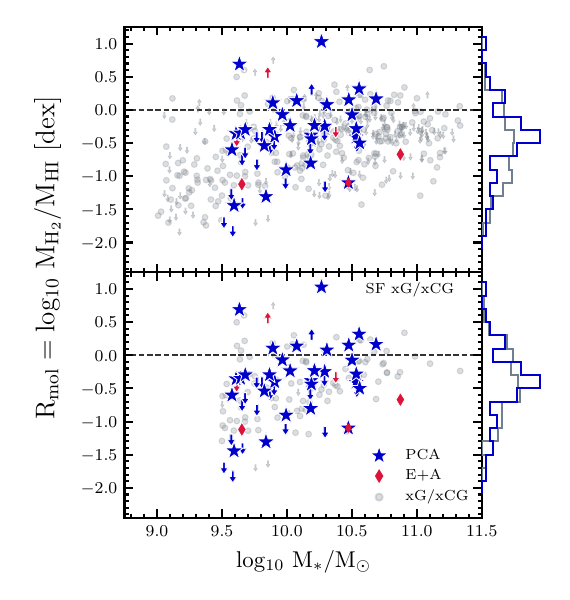}
    \caption{ {\textit{Top:} The ratio of molecular to atomic gas, $\rm R_{mol}$, as a function of stellar mass for both PSBs (blue and red) and all galaxies in both xGASS and xCOLD GASS (grey) detected in either CO or H$\text{\sc{i}}$. When CO is detected but not HI, an upward facing arrow denotes a lower limit on the ratio, while an upper limit is shown with a downwards arrow when HI is detected but not CO. Non-detections in both gas phases are omitted from this figure. Equal gas content in both gas phases is denoted as a horizontal dashed line at $\rm R_{mol}=0$. To the right of the plot are density histograms of $\rm R_{mol}$ for both EMBERS PSBs and xGASS/xCOLD GASS. \textit{Bottom:} The same as the top panel but comparing PSBs to just star forming galaxies in xGASS/xCOLD GASS with log$_{10}~$M$_{\ast}$/M$_{\odot}>9.5$. $\rm R_{mol}$ spans $\sim 3$ dex in both our PSBs and the comparison sample, with some galaxies in both samples hosting larger molecular gas reservoirs than atomic. When compared to their progenitor population of star-forming galaxies in the same mass range, PSBs have similar $\rm R_{mol}$ distributions to a comparison sample, indicating that inefficient H$\text{\sc{i}}$-to-H$_2$ conversion is not the cause of quenching. }}

    \label{fig:molecular-atomic}
\end{figure}

\begin{equation}
    \rm R_{mol} = log_{10} \frac{M_{H_2}}{M_{H\text{\sc{i}}}} .
\end{equation}

\noindent {If molecular gas is used up in star formation, such as in a recent starburst, and if the extended H$\text{\sc{i}}$ reservoir does not efficiently replenish the H$_{2}$ reservoir, we should expect $\rm R_{mol}$ to drop compared to normal galaxies. An observed bottleneck in $\rm R_{mol}$ could be what initiated the quenching of a starburst, such as what we see in PSBs. Conversely, if $\rm R_{mol}$ is elevated compared to normal galaxies, as is seen in galaxies with the highest specific star formation rates (sSFR; see \citeauthor{saintongeMolecularAtomicGas2016} \citeyear{saintongeMolecularAtomicGas2016}), it is more likely that quenching is initiated due to the conditions of the molecular gas and not the fault of the conversion of the atomic gas phase to the molecular gas phase.}

{The top panel of Figure \ref{fig:molecular-atomic} shows the values of $\rm R_{mol}$ for all galaxies that were observed in both xGASS and xCOLD GASS in grey as a function of stellar mass. EMBERS PSBs are shown as coloured points when both measurements are available. When a galaxy is detected only in H$\text{\sc{i}}$, we denote the upper limit as a downward arrow. Similarly, when a galaxy is detected only in CO, we denote the lower limit as an upward arrow. When both measurements in H$\text{\sc{i}}$ and CO are not securely detected, we omit the galaxy since $\rm R_{mol}$ is unconstrained (eight EMBERS PSBs are not detected in either gas phase). The right hand axis of Figure \ref{fig:molecular-atomic} shows the density histogram of all points included in the figure for both the xGASS/xCOLD GASS comparison and EMBERS.}
{In the comparison sample (xGASS/xCOLD GASS), $\rm R_{mol}$ spans nearly 3 orders of magnitude, with values typically less than 0 indicating more atomic gas than molecular, as found by \citet{saintongeMolecularAtomicGas2016} and \citet{catinella_xgass_2018}. The behaviour of EMBERS PSBs is similar, with galaxies ranging from 50 times more atomic gas than molecular to galaxies with larger molecular gas reservoirs than atomic. EMB12 is an outlier in this way, with $\sim10$x more molecular gas than atomic. However, although there are PSBs across the entire dynamic range of the comparison sample, the typical PSB is centred around a median of $\rm R_{mol}=-0.39$ compared with a median $\rm R_{mol}=-0.61$ in the comparison sample. Moreover, the comparison galaxies are less concentrated around their median value compared to our PSBs. A Kolmogorov-Smirnov (K-S) test gives a p-value of 0.004, highlighting the difference between the two underlying distributions.}

{As many of our PSBs show signs of elevated $\rm R_{mol}$ compared to galaxies not in the process of quenching, and as a population are \textit{not} systematically suppressed in $\rm R_{mol}$, we might say that it is not molecular gas depletion followed by inefficient refueling of the molecular gas reservoir via extended atomic gas that leads to quenching in the majority of our PSBs. However, there are subtleties associated with our interpretation of the upper panel in Figure \ref{fig:molecular-atomic}. Although we see a 0.2 dex enhancement compared to all of xGASS/xCOLD GASS, a more appropriate comparison sample is star forming galaxies within the same mass regime, as PSBs must have originated from a star forming progenitor galaxy. Moreover, many of the low $\rm R_{mol}$ values in the comparison sample are from galaxies with log$_{10}~$M$_{\ast}$/M$_{\odot}<9.5$, which are absent in EMBERS.}

{In order to have a fairer comparison between PSBs and their progenitor population, we limit the xGASS/xCOLD GASS sample to just star forming galaxies defined  using the same BPT classification from Section \ref{sec:gas fractions} and require log$_{10}~$M$_{\ast}$/M$_{\odot}>9.5$. The bottom panel of Figure \ref{fig:molecular-atomic} shows our PSBs compared to just the SF subset of comparison galaxies, where we can see that the distributions are largely similar (in contrast to the upper panel). The median of the SF comparison sample is $\rm R_{mol} = -0.43$ compared to $\rm R_{mol}=-0.39$ for our PSBs, while a K-S test results in a p-value of 0.6 such that we cannot say they are drawn from different distributions. We then conclude that the molecular-to-atomic gas ratio in PSBs is consistent with that of normal star forming galaxies and is not the reason behind the quenching of star formation in EMBERS PSBs.}

{We have shown that EMBERS PSBs as a population are deficient in molecular gas and that their molecular-to-atomic gas ratio, $\rm R_{mol}$, is not atypical of normal star forming galaxies, but that individual galaxies vary considerably in both quantities. We now look to quantify the individual relative gas content of our PSBs on a galaxy-by-galaxy basis compared to our bespoke xCOLD GASS control sample.}

\begin{figure*}
  \centering
  \includegraphics[width=0.45\textwidth]{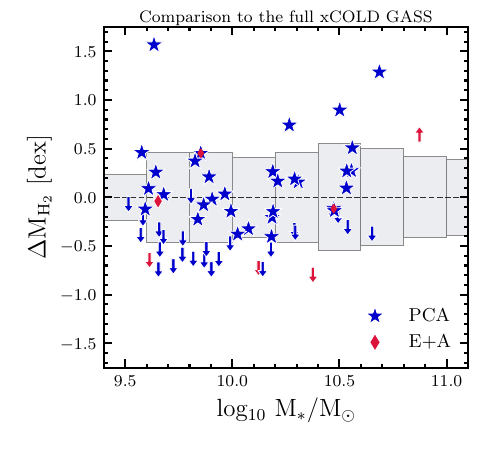}\hspace{0.0\textwidth}
  \includegraphics[width=0.45\textwidth]{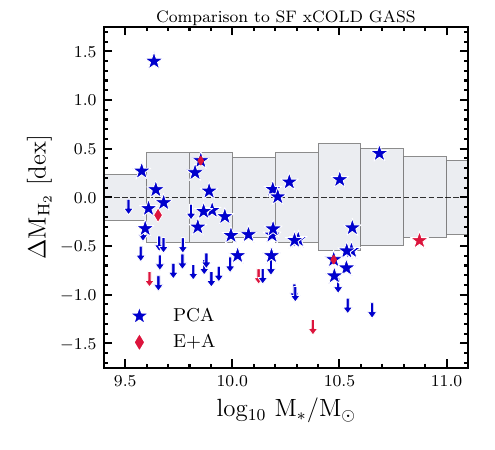}
  \caption{\textit{Left:} Offset from mass-matched controls drawn from all of xCOLD GASS as a function of stellar mass. The dashed black line corresponds to a galaxy as `gas-normal' relative to the entire xCOLD GASS population. Non-detections in CO(1--0) are denoted with downwards arrows. The shaded area corresponds to the $\pm 1 \sigma$ spread of $\Delta$MH$_2$ for galaxies in the xCOLD GASS comparison sample. \textit{Right:} the same but for a BPT selected SF control subset of xCOLD GASS as a control sample. This figure highlights the diversity of gas properties seen in our sample but also that the majority of PSBs in the sample when compared the star forming progenitors, are significantly gas poor by $\sim 0.5-1.3$ dex. The detection fraction for matches to EMB55 is $<50$ per cent what controls are drawn from the full xCOLD GASS sample, and so we denote it as a lower limit in the left panel.  }
  \label{fig:deltas}
\end{figure*}

\subsection{Gas mass offsets from controls}
\label{sec:deltas}
\subsubsection{Molecular gas offsets}

Having established that the typical PSB is gas-poor compared to comparison galaxies (as shown by Figure \ref{fig:fractions}) but that some PSBs host large molecular gas reservoirs (see Figure \ref{fig:gass distributions}), we now compare individual galaxies to controls. We compute an offset from the control sample, $\Delta \rm M_{H_2}$, to quantify precisely how different the gas properties of our PSBs are compared to non-PSBs of the same stellar mass. The $\Delta \rm M_{H_2}$ diagnostic allows us to capture the relative excess, or depletion, of a galaxy given its stellar mass compared to galaxies within the same regime drawn from our control pool of xCOLD GASS. 

To compute $\Delta \rm M_{H_2}$, we populate a control sample, drawn from xCOLD GASS, for each individual PSB such that the stellar masses match within 0.1 dex of the PSB. In doing so, we are able to match our PSBs to between 30-70 control galaxies each. We then compute the offset ($\Delta \rm M_{H_2}$) as the difference between the measured molecular gas mass for the PSB and the \textit{median} molecular gas mass of the matched controls. For our offset to be robust, we require that the detection fraction in the control pool to be $>50\%$ (i.e., the computed median is a CO detection in xCOLD GASS). Only one galaxy, EMB55, fails this requirement in matching due to its high stellar mass and the larger gas-poor population in xCOLD GASS in this regime. As such, we denote it as an upward arrow \textit{lower} limit in the left panel of Figure \ref{fig:deltas}.

The left panel of Figure \ref{fig:deltas} shows $\Delta \rm M_{H_2}$ as a function of stellar mass for all PSBs, with non-detections denoted as downward arrows using galaxies drawn from the entirety of xCOLD GASS as controls. We also compute $\Delta \rm M_{H_2}$ for each galaxy in the xCOLD GASS sample and overlay the typical $1\sigma$ deviation as the background shaded region in bins of 0.2 dex in stellar mass to show the expected scatter in non-PSB galaxies. The horizontal dashed line corresponds to an offset of 0, indicating that a galaxy is gas-normal for its stellar mass. The majority of PSBs fall within the spread around 0 when compared to controls drawn from all of xCOLD GASS, suggesting that (on average) the PSBs might be considered gas-typical for their stellar mass. Nonetheless, there exists a population ($\sim15$ galaxies) of galaxies that have not been detected in CO down to an upper limit in gas mass offset of $3-8\times $ gas deficient. Conversely, a handful of PSBs ($\sim4-5$ galaxies) host significant molecular gas reservoirs relative to similar galaxies in xCOLD GASS. We highlight EMB48 and EMB85 as being $\sim20$x and $\sim 40$x gas-rich compared to their xCOLD GASS controls. In summary, although there are outliers in both directions, the majority of PSBs fall within $1\sigma$ of non-PSB galaxies that include both SF and quenched, a result which holds for both PCA-selected PSBs and E+A PSBs. 

The progenitor population of our PSBs are presumably SF galaxies as the PSBs are in the process of quenching. As such, the SF subset of xCOLD GASS makes a more suitable control sample with which to compute offsets from. We apply the same BPT cut used in Section \ref{sec:gas fractions} to make a distinction between the entire set of controls and just the SF subset. Using the SF control pool, we compute $\Delta \rm M_{H_2}$ with between 11 and 40 matches per PSB (except EMB55). All matched control samples have robust detection fractions, but EMB55 only achieves 3 SF matches due to its high mass and so we expand the mass match limits for EMB55 to 0.15 resulting in 10 matches. The right-hand panel of Figure \ref{fig:deltas} shows $\Delta \rm M_{H_2}$ as a function of stellar mass for PSBs matched to SF xCOLD GASS. The shaded region is the $1\sigma$ spread in $\Delta \rm M_{H_2}$ computed for the SF subset (which is slightly tighter than for all of xCOLD GASS). 

Only one galaxy, EMB85, remains gas-rich (above $1\sigma$ of SF xCOLD GASS) compared to SF xCOLD GASS galaxies, retaining $\sim25$x more gas than its SF counterparts. Less than half of PSBs (27/61) host gas reservoirs \textit{within} the $1\sigma$ range of SF xCOLD GASS galaxies. The remaining 34 PSBs are gas-poor (below $1\sigma$ of SF xCOLD GASS). We then conclude that most PSBs are gas-poor relative to typical star-forming galaxies. That being said, the diversity in gas properties across the sample seen in Figure \ref{fig:gass distributions} persists here in Figure \ref{fig:deltas} as gas depletion is not ubiquitous for all PSBs and $\Delta \rm M_{H_2}$ spans almost 2 dex (excluding EMB85, which is a further order of magnitude enhanced). 

Of the four galaxies with beam covering fractions $<1$ mentioned in Section \ref{sec:methods}, three of them have positive $\Delta \rm M_{H_2}$ offsets with respect to the SF subset of xCOLD GASS. These galaxies are EMB09, EMB48, and EMB62, and the inclusion of an aperture correction to their CO observations would only make their respective $\Delta \rm M_{H_2}$ more positive and the conclusions about these galaxies would not change. EMB55 has a negative $\Delta \rm M_{H_2}=-0.44$ so that an aperture correction could move this galaxy from the ``gas-poor'' to ``gas-normal'' regime. The maximum aperture correction done by the xCOLD GASS survey is 2.4x; EMB55 is quite extended, and so if we assume that the maximum correction is required, this galaxy would move from relatively gas poor to gas-normal (upwards by 0.38 dex). 

In the molecular gas phase, we find a comparable but more pronounced result compared to \citetalias{ellisonLowRedshiftPoststarburst2025a}; a larger fraction of EMBERS PSBs are depleted in molecular gas than atomic gas when compared to SF xCOLD GASS, but both gas phases exhibit diversity. Access to both the atomic and molecular gas phases allows us to assess the total gas content of our PSBs in tandem for the first time.

\subsubsection{Molecular + atomic gas mass offsets from controls}

\begin{figure}
\centering
    \includegraphics[width=0.5\textwidth]{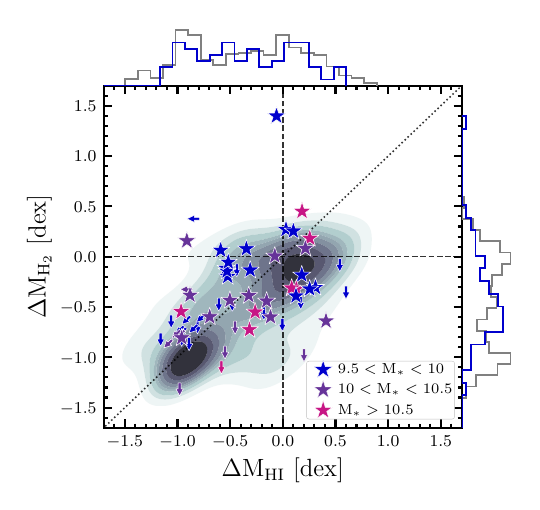}
    \caption{Gas mass offset from the BPT selected star-forming control sample of xCOLD GASS and xGASS for molecular and atomic gas for our PSBs. The PSB points are coloured in stellar mass bins of 0.5 with low mass PSBs in blue, intermediate mass PSBs in purple, and high mass PSBs in violet. If a galaxy is detected only in HI, the limit is denoted as a downward arrow, if a galaxy is only detected in CO the limit is denoted as a left-pointing arrow, and non-detections in both gas phases are denoted as 45 degree angles down and left. The horizontal and vertical dashed lines correspond to `gas-normal' in a given gas phase compared to the SF population of controls while the diagonal dotted line is the 1-to-1 line where a galaxy is equally depleted in both gas phases. In the background is the gas mass offsets for the combined comparison sample relative to the SF subset. Contours enclose a percentage of the entire population of controls in steps of 10$\%$ (i.e., the outer contour encloses 100 per cent of the xCOLD GASS/xGASS sample). Along the right and top edges of the plot are projections of the variable along each axis for both the PSBs (blue) and controls (grey). This figures highlights how the majority of the PSBs in our sample exist in a transitionary state in cold gas between star-forming and quenched, a regime which is largely not populated by the xGASS and xCOLD GASS samples. 
    }
    \label{fig:delta-delta}

\end{figure}

\citetalias{ellisonLowRedshiftPoststarburst2025a} computed a similar offset as we have done in Section \ref{sec:deltas} but for the H$\text{\sc{i}}$ content of our PSBs relative to xGASS. By combining the offset in atomic gas with that of molecular gas presented here, we can evaluate the relative multiphase gas content of our PSBs compared to controls. The two offsets, $\Delta \rm M_{HI}$ and $\Delta \rm M_{H_2}$, computed with respect to BPT selected SF galaxies in each control pool (as selected in Section \ref{sec:gas fractions}), are shown in Figure \ref{fig:delta-delta} with the molecular gas offset as a function of the atomic gas offset. The PSBs are shown as coloured markers with horizontal arrows for atomic gas non-detections, vertical arrows for molecular gas non-detections, and angled arrows when both phases are not detected. Horizontal and vertical dashed lines correspond to ``gas-normal'' in a given phase while the diagonal dotted line is the 1-to-1 line corresponding to an equal offset in both gas phases. The grey contours in the background are the same quantities computed for galaxies found in both xCOLD GASS and xGASS matched against the SF galaxies in each sample, respectively. Each contour line encloses 10 per cent of the total distribution of comparison galaxies. The top and left edges of the figure show the distributions of gas mass offsets in either phase for both the PSBs and comparison galaxies. 

Figure \ref{fig:delta-delta} highlights the diversity in gas properties of our PSBs. Each quadrant is occupied by at least five PSBs, demonstrating that PSBs can be gas-rich in both phases (top right quadrant), just HI-rich (bottom right), just H$_2$-rich (top left), or gas-poor (bottom left). That being said, the majority of PSBs are preferentially poor in both phases, with 38/58 (all PSBs with both phases) galaxies occupying the bottom left quadrant. Most PSBs fall within the underlying distribution of the comparison sample (grey contours) of normal galaxies; only four galaxies fall outside of the grey contours. EMB85 is once again a considerable outlier as it is HI-normal with a 25x excess of molecular gas, whereas EMB12 and EMB69 are highly H$_2$ rich given their H$\text{\sc{i}}$ offset and EMB48 is slightly more gas-rich than the galaxies in xGASS/xCOLD GASS.

In order to understand the connection between gas depletion and the quenching of star formation, we begin by characterizing the behaviour of the xGASS and xCOLD GASS samples. We find that both of the gas mass offsets ($\Delta \rm M_{H\text{\sc{i}}}$ and $\Delta \rm M_{H_2}$) plotted in Figure \ref{fig:delta-delta} for the \textit{comparison} sample correlate strongly with the star formation offset from the main sequence ($\Delta$SFR) with SFRs taken from the MPA/JHU catalog \citep{brinchmann_2004}. With $\Delta \rm M_{H\text{\sc{i}}}$ we find a Pearson correlation coefficient of $r=0.56$ (p$~\ll 10^{-4}$) while $\Delta \rm M_{H_2}$ correlates with $\Delta$SFR with coefficient $r=0.79$ (p$~\ll 10^{-4}$), indicating strong correlations, especially with molecular gas mass offsets. Additionally, $\Delta \rm M_{H\text{\sc{i}}}$ and $\Delta \rm M_{H_2}$ for the comparison sample correlate strongly with each other giving $r=0.55$ (p$~\ll 10^{-4}$). These results mean that as a galaxy departs from the main sequence, both gas phases deplete in a coherent way through the green valley and into the quenched population, or more likely, the gas depletion leads to quenching. 

The correlations we find between $\Delta \rm M_{H\text{\sc{i}}}$, $\Delta \rm M_{H_2}$, and $\Delta \rm SFR$ are consistent with results presented by \citet{saintaonge_catinella_annurev_2022} where, along the main sequence ($\Delta\rm SFR \sim 0$) we see gas mass offsets of [$\Delta \rm M_{H\text{\sc{i}}}$,~$\Delta \rm M_{H_2}$]~$\sim[0,~0]$, which smoothly decrease to a similar buildup of quenched galaxies ($\Delta\rm SFR < -1 $) at $[\Delta \rm M_{H\text{\sc{i}}}$,~$\Delta \rm M_{H_2}]$~$\sim[-1.1,~-1.2]$.\footnote{We note here that the first locus around $[\Delta \rm M_{H\text{\sc{i}}}$,~$\Delta \rm M_{H_2}]$~$\sim[0,~0]$ for the comparison galaxies is slightly offset from the midpoint of Figure \ref{fig:delta-delta}; the distributions of offsets for the SF control population \textit{are} centred on 0 in both gas phases and are relatively symmetric around the origin. The slight offset comes from galaxies that are rich in H$\text{\sc{i}}$ (or in some cases $\rm H_2$) but do not have emission lines consistent with the \citet{kauffmann_bpt_2003} SF cut (i.e., AGN dominated emission lines). } Although the quantities in Figure \ref{fig:delta-delta} are slightly different than the ones computed by  \citet{saintaonge_catinella_annurev_2022} (offset from all SF galaxies and not just galaxies on the SF main sequence), the same qualitative picture holds: star-forming galaxies inhabit a locus near the origin, and quenched galaxies inhabit a locus at around $[\Delta \rm M_{H\text{\sc{i}}}$,~$\Delta \rm M_{H_2}]$~$\sim[-1,~ -1]$. The two peaks in the distribution can be clearly seen in the contours of Figure \ref{fig:delta-delta} and the grey histograms on the top and right of the plot, where a bimodality of gas mass offsets in both the atomic and molecular phases is found.

Having established that the background histogram of Figure \ref{fig:delta-delta} displays an evolutionary pathway in which xGASS/xCOLD GASS galaxies are depleted in both H$\text{\sc{i}}$ and $\rm H_2$ as they quench in star formation, we find that our PSBs are not found preferentially in either the gas-rich or gas-poor locus, but instead are most commonly in an intermediate regime. The majority of PSBs therefore exist in a transition state between the star-forming locus of galaxies which host large cold gas reservoirs and the quenched locus of gas-depleted galaxies. We also find that the correlation between $\Delta \rm M_{H\text{\sc{i}}}$ and $\Delta \rm M_{H_2}$ in our PSBs is weaker, but still significant, than the $r=0.55$ found for the comparison sample with $r=0.41$ (p=0.002).

In the atomic phase, the transition between gas-rich and quenched galaxies in xGASS is smoother, with a weaker bimodality than that seen for molecular gas. Even so, our PSBs are not drawn from the same distribution, showing an excess of intermediate $\Delta \rm M_{H\text{\sc{i}}}$ values relative to the comparison sample. A Kolmogorov-Smirnov (K-S) test demonstrates this difference with a p-value of 0.012. The contrast between our PSBs and xCOLD GASS (i.e., in their molecular gas properties) is even more evident. xCOLD GASS galaxies are highly bimodal in $\Delta \rm M_{H_2}$, whereas the distribution of our PSBs is single-peaked and continuous, with the majority of galaxies occupying the space \textit{between} the two peaks in xCOLD GASS (see the right-hand marginal histogram in Figure \ref{fig:delta-delta}). A K-S test of the two $\Delta \rm M_{H_2}$ distributions gives a p-value of $8.1\times 10^{-6}$, highlighting the significance of the difference between molecular gas mass offset distributions between our PSBs and xCOLD GASS. This is consistent with the picture that PSBs are actively transitioning from one locus in $\Delta \rm M_{H\text{\sc{i}}} - \Delta \rm M_{H_2}$ space to the other and that this transition is predominantly driven by the molecular gas content.

Incorporating SFRs into our analysis, as well as characterizing the burst properties (strength and timing) would provide additional insight into the quenching of PSBs.  Unfortunately, estimating SFRs for PSBs is notoriously difficult due to the timescales associated with and potential lack of typical SFR tracers \citep{yesuf_2017, frenchStateMolecularGas2023c, wild_ifsfr_2025}.  However, in a future work (Rasmussen et al. in prep) we will use detailed star formation histories to constrain these properties.  Nonetheless, even without reliable measurements of SFRs, we have placed PSBs in a region of transition from star-forming to quenched using only their gas mass offsets from normal SF galaxies, suggesting that consumption or evacuation of gas is partly to blame for their quenching. Even still, many of our PSBs have either atomic or molecular gas (or both) at comparable rates to SF galaxies and, despite this, quenching has still been initiated in these galaxies. 

In summary, there is a diversity in gas properties in both the atomic and molecular gas seen in EMBERS PSBs with at least 5 galaxies occupying each quadrant of Figure \ref{fig:delta-delta}. That being said, the majority of galaxies in the EMBERS sample show depleted gas reservoirs in both phases that are consistent with masses intermediate to those of gas-rich, star-forming galaxies and gas-poor, quenched galaxies. We now look to investigate how robust our results are with respect to assumptions made in Section \ref{sec:methods}. 
\section{Discussion}
\label{sec:discussion}

\subsection[The effects of the CO-to-H2 conversion factor]{The effects of the CO-to-H\textsubscript{2} conversion factor}
\label{sec:alpha}

A major assumption made to convert CO luminosities to the inferred gas mass is our choice of conversion factor $\alpha_{\rm CO}$. Having concluded that the majority of PSBs are gas-poor compared to star-forming galaxies, we investigate the impact of our initial constant conversion factor assumption. Contemporary prescriptions for a variable conversion factor typically require a gas-phase metallicity term combined with a metric of the star formation or the conditions of the ISM at either the unresolved or resolved scale \citep{bolatto_alpha_2013,sandstrom_2013,accurso_alpha_2017,chiang_2024,schinnerer_leroy_2024rev}. As our observations are unresolved, the kpc-scale prescriptions are unsuitable, but we investigate the qualitative effect of their application on both our PSBs and the comparison sample to mitigate the effects of the conversion factor assumption on our overall conclusions.

Accounting for variations in $^{12}\rm CO$ opacity at the kpc scale is captured with either a starburst emissivity term like that of \citet{chiang_2024} or a dependence on the velocity dispersion of the molecular gas as done by \citet{teng_2024}. The starburst emissivity term is dependent on the stellar mass surface density, and can only provide a conversion factor lower than that used here. The CO emission in PSBs is more compact than that of normal galaxies (see \citeauthor{smercinaFallResolvingMolecular2022} \citeyear{smercinaFallResolvingMolecular2022}; \citeauthor{otterResolvedMolecularGas2022} \citeyear{otterResolvedMolecularGas2022}) \textit{and} PSBs typically have smaller sizes than normal galaxies at a fixed stellar mass for fibre-selected or central PSBs, of which all of our PSBs are \citep{chen_2022,cheng_2024}. Thus, the observed molecular gas in PSBs is more likely to be co-spatial with regions of higher stellar mass surface density compared to xCOLD GASS comparison galaxies at a fixed stellar mass. Applying a starburst emissivity term to both the PSBs and xCOLD GASS strengthens the depletion that we observe in Figures \ref{fig:fractions} and \ref{fig:deltas} as the value of $\alpha_{\rm CO}$ would decrease at higher rates for PSBs than for the controls with the prescription from \citet{chiang_2024}. 

The majority of our PSBs show morphological signatures consistent with a recent galaxy merger (see \citeauthor{wilkinson_2022} \citeyear{wilkinson_2022} for discussion on PSBs as a population). Galaxy mergers have been shown to have elevated cloud-scale molecular gas velocity dispersions \citep{he_2023}. Additionally, resolved studies have shown high sub-kpc scale velocity dispersions caused by turbulence in PSBs \citep{smercinaFallResolvingMolecular2022} and so it is likely that PSBs have elevated dispersions relative to xCOLD GASS controls. The velocity dispersion dependent prescription from \citet{teng_2024} smoothly decreases with increasing velocity dispersions and, as such, the decrements seen in Figures \ref{fig:fractions} and \ref{fig:deltas} are once again \textit{underestimated} by using a constant $\alpha_{\rm CO}$. 

The one component relating CO emission to total bulk molecular gas that would decrease the deficiencies in gas compared to xCOLD GASS (i.e., make them more gas-normal) that we see is the gas-phase metallicity. As many of our PSBs do not have nebular emission lines, we are unable to measure the gas-phase metallicity directly. We infer a metallicity for each EMBERS PSB by assuming that the mass-metallicity relation (MZR) holds \citep{tremonti_2004,kewley_2008}, an assumption which is likely conservative. \citet{leung_2024} showed that the stellar metallicities for PSBs are heavily enriched in the time since the starburst. Additionally, although stellar metallicities are not an exact predictor of gas-phase metallicities, emission line analysis by \citet{boardman_2024} has shown that the gas in PSB regions is not metal poor, meaning that the MZR may underestimate PSB metallicities.

We compute metallicities by matching each PSB in stellar mass within 0.1 dex and redshift within 0.01 dex to all galaxies from SDSS DR7, obtaining between 79 and 2837 matches per PSB. Using metallicities calibrated via the method of \citet{pettini_2004} (hereafter PP04) with [OIII] and [NII] emission lines, we compute the median metallicity of the matched (non-PSB) control pool. This median metallicity of the $\rm M_{\ast}$ and $z$-matched sample is then assumed to be typical for the PSB. We then apply the prescription from \citet{accurso_alpha_2017}:

\begin{figure}
  \centering
  \includegraphics[width=0.45\textwidth]{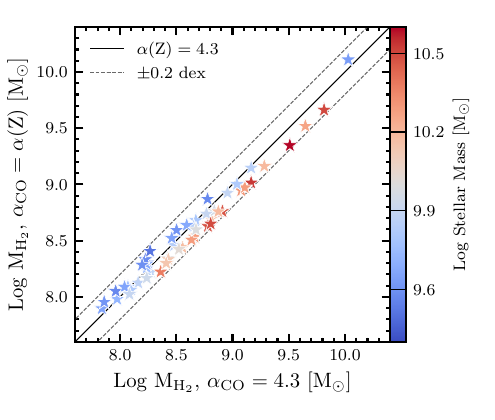}
  \caption{The molecular gas mass inferred by a metallicity dependent CO-to-H$_2$ conversion factor, $\alpha_{\rm CO}(\rm Z)$ based on \citet{accurso_alpha_2017} as a function of the molecular gas mass inferred with a constant $\alpha_{\rm CO}=4.3~\rm M_{H_{2}}/L'_{CO}$. All points are coloured by the galaxy's total stellar mass. The solid black diagonal line corresponds to $\alpha_{\rm CO}(\rm Z)=4.3$ while the black dashed lines show an increase/decrease of 0.2 dex from the values quoted in Table \ref{tab:psb_co10}. This figure demonstrates how the assumptions leading to our computation of $\rm log ~M_{H_2}$ do not affect the molecular gas deficiency that we observe. }
  \label{fig:change in alphas}
\end{figure}

\begin{equation}
    \rm log ~ \alpha_{\rm CO} = 14.752~-~1.623[12+log(O/H)]~+~0.062~log~\Delta SFR,
    \label{eq:accurso}
\end{equation}
\noindent where $\rm \Delta SFR$ is the star formation offset from the main sequence and 12+log(O/H) is the metallicity in PP04 units. The contributions from the second term in Equation \ref{eq:accurso} are small, and our PSBs are actively quenching in SFR. Therefore, since we are most interested in exploring scenarios in which we might have underestimated $\rm \alpha_{\rm CO}$ (i.e. underestimated the gas content of PSBs), we omit the second term in Equation \ref{eq:accurso}. 

Figure \ref{fig:change in alphas} shows the molecular gas mass of our PSBs using a conversion factor from Equation \ref{eq:accurso} versus our fiducial H$_2$ masses quoted in Table \ref{tab:psb_co10}, coloured by stellar mass. It is clear from Figure \ref{fig:change in alphas} that even in the lowest metallicity (low mass due to MZR) cases, the inferred molecular gas mass never increases by more than 0.2 dex and that it is more likely to decrease. {Although we choose to apply the \citet{accurso_alpha_2017} metallicity term, similar relations from \citet{hunt_2020} and \citet{schinnerer_leroy_2024rev} have very comparable dependencies, resulting in $\alpha_{\rm CO}$ that deviates by no more than 0.2 dex from our fiducial constant $\alpha_{\rm CO}$. Thus, the maximal changes seen in Figure \ref{fig:change in alphas} are largely not sensitive to our choice of prescription. }



A deficiency in molecular gas mass seen when computing offsets from a control pool ($\Delta \rm MH_{2}$) is likely invariant to changes in the conversion factor that arise from galactic parameters that depend primarily on stellar mass, such as metallicity, due to our matching to a fixed stellar mass. This means that the deficiency demonstrated by Figure \ref{fig:deltas} remains largely intact regardless of assumptions with $\rm \alpha_{\rm CO}$. 

By testing the current literature prescriptions of $\rm \alpha_{\rm CO}$, we find that in the majority of cases the depletion of molecular gas that we observe would strengthen. {Our derived molecular gas masses are not sensitive to more than 0.15 dex changes from a metallicity based prescription \citep{accurso_alpha_2017,hunt_2020,schinnerer_leroy_2024rev}, whereas it is likely that the inclusion of an emissivity term \citep{bolatto_alpha_2013,chiang_2024} or a CO velocity dispersion term \citep{teng_2024} would increase the magnitude of the depletions seen in Figure \ref{fig:fractions} and \ref{fig:deltas}.} Thus, we can say with some certainty that the majority of low-redshift PSBs are gas-poor compared to their star-forming progenitors. In the cases where CO is not detected in EMBERS PSBs, it is of use to try and obtain a detection at the $5\sigma$ level by stacking spectra of multiple galaxies. 

\subsection{Stacking of non-detections}
\label{sec:stacking}
We stack spectra from multiple PSBs to get a better idea of how gas-poor our CO non-detections are. To do this, we split all non-detected PSBs with CO Source of 4 in Table \ref{tab:psb_co10} (i.e., observed in our IRAM 30m observing program) into bins of 0.25 dex in stellar mass. This results in four bins with nine PSBs in each low-mass bin and 3 and 4 PSBs in the two higher mass bins, respectively. Additionally, we stack \textit{all} low mass non-detections (log$_{10}$ M$_{\ast}$/M$_{\odot}<10$) for completeness. We centre each emission line on the galaxy's spectroscopic redshift and average all scans from all galaxies in each bin. We compute emission line information and spectrum noise in the same fashion for the stacked spectrum as done in Section \ref{sec:observations} and the results of stacking can be found in Table \ref{tab:stacks}. The molecular gas masses and gas fractions are computed by taking the median stellar mass and redshift in each of the stacked bins combined with Equations \ref{eq:luminosity} and \ref{eq:alpha L}.

\begin{table*}
    \centering
    \caption{Measured quantities from stacked non-detections in bins of stellar mass}
    \setlength{\tabcolsep}{9pt}

    \begin{tabular}{cccccccc}
    \hline  
         PSBs in Stack& Stellar Mass Range  & $\int \mathrm{F}_v \mathrm{d}\nu$ & RMS& S/N&L$_{\text{CO}}$ & log$_{10}~$MH$_2$ & $f_{\rm H_2}$\\
         (EMBERS $\#$) &(M$_{\sun}$) &(mK km s$^{-1}$)& (mK)&  & ($10^7$ K km s$^{-1}$ pc$^2$) & (M$_{\sun}$) & \\
         \hline
         
         21, 36, 38, 45, 49, 51, 61, 70, 71  &9.5 -- 9.75   &0.023&0.18  & 3.01& $<$0.76& $<$7.52 &$<$0.0073 \\
         07, 08, 10, 18, 23, 26, 28, 52, 74  &9.75 -- 10.0  &0.054&0.24  & 3.62& $<$0.99& $<$7.63 &$<$0.0058 \\
         00, 66, 68                          &10.0 -- 10.25 &0.13&0.55   & 3.12& $<$3.65& $<$8.20 &$<$0.011  \\
         39, 44, 59, 64                      &10.25 -- 10.5 &0.44&0.55   & 7.86& 12.9   & 8.74    &0.023   \\
         Rows 1+2                            &9.5 -- 10.0   &0.032&0.13  & 5.31& 0.59   & 7.41    &0.0046  \\
         
        \hline 
    \end{tabular}
    {\raggedright  \textit{Note.} --- Columns are:
    1 -- PSBs involved in spectral stack;
    2 -- Stellar mass range that non-detections are drawn from;
    3 -- Integrated CO flux of stacked spectrum;
    4 -- RMS of stacked spectrum;
    5 -- S/N of stacked spectrum;
    5 -- CO line luminosity of stacked spectrum;
    6 -- Log of molecular gas mass inferred from CO emission using median mass and redshift of stacked sources;
    7 -- Gas fraction of median galaxy in stack.
   }
    \label{tab:stacks}
\end{table*}

\begin{figure}
 \includegraphics[width=0.45\textwidth]{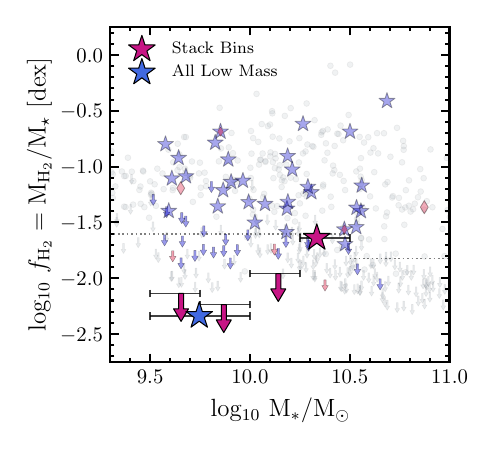}

  \caption{Cutout of the right panel of Figure \ref{fig:gass distributions} with the results of the stacking of all non-detections in bins of stellar mass included. In violet, represented by the downwards arrows for non-detections and a star for a detection show results of all stacked non-detections in a given bin of 0.25 dex in stellar mass. We compute gas masses using the median mass and redshift of galaxies in the bin. The blue star is the result of stacking every galaxy with $9.5 <~$log$_{10}$$~\rm M_{\ast}/M_{\odot}$$<10$. The stellar mass ranges of each point are shown with horizontal solid lines. The reference points from Figure \ref{fig:gass distributions} have been dimmed for ease of comprehension. The typical low-mass PSB hosts a molecular gas reservoir of less than 0.46 per cent of its total stellar mass. The depletion that we see in molecular gas for these galaxies in Figure \ref{fig:deltas} would decrease by up to 0.5 dex given enough integration time to secure detections. }
  \label{fig:result of stacks}
\end{figure}

Figure \ref{fig:result of stacks} places the inferred molecular gas mass for the median PSB in each stacked bin onto the right panel of Figure \ref{fig:gass distributions} to highlight the gas fraction expected of the PSBs in each bin. Non-detections in a given stack are denoted with a downward arrow, whereas a detected stack is shown with a coloured star. The range in stellar mass of each point is denoted with a horizontal bar. The three lower mass bin stacks are not detected at the $5\sigma$ level despite considerable equivalent integration times. For the first three bins, the spectral stacks equate to 60.3, 40.1, and 12.0 hours of IRAM 30m time, in order of increasing stellar mass respectively. Given the depth of the stacked observations, we achieve gas fraction upper limits of 0.73, 0.58, and 1.1 per cent for the three lower mass stacks, all of which decrease our individual upper limits of the galaxies occupying each bin by between 0.2 and 0.5 dex in gas fraction. The highest stellar mass bin is detected at the $5\sigma$ level corresponding to a gas fraction for these galaxies of $\sim 2.3$ per cent, right below our detection threshold. 

When we stack all 18 galaxies with log$_{10}$ M$_{\ast}$/M$_{\odot}<10$, we achieve a detection at the $5\sigma$ level. The total integration time included in this stack exceeds 100 hours and results in a gas fraction for the median galaxy (in mass/redshift) in the subsample of 0.46 percent. If we take this to mean that all galaxies in the stack have a gas fraction on the order of this, the depletion of the lower stellar mass regime in Figure \ref{fig:fractions} would decrease by up to 1.0 dex. Additionally, on a galaxy-by-galaxy basis, the depletion in Figure \ref{fig:deltas} would decrease by between $\sim 0.4-1.1$ dex. Thus, we conclude that there exists a population of PSBs that are strongly depleted in molecular gas that exists preferentially in the low stellar mass regime.

\subsection{Comparison to previous works}
Global molecular gas has been measured in PSB galaxies in a number of prior works. \citet{frenchDISCOVERYLARGEMOLECULAR2015g} measured CO in 32 nearby (0.01 $<z<$ 0.12) E+A PSBs and detected molecular gas in 53 per cent of cases. The majority of galaxies in the sample have $\rm log_{10} ~M_{\ast}/M_{\odot}>10$ with molecular gas masses ranging from $8.6<\rm log_{10} ~M_{H_{2}}/M_{\odot}<9.8$. They conclude that a rapid reduction in SFR is not always accompanied by a dearth of molecular gas. Similarly, \citet{rowlands_psbs_2015} observed molecular gas in 11 PCA selected PSBs and detected gas in 91 per cent of their PSBs. 8/11 of the PSBs in \citet{rowlands_psbs_2015} have $\rm log_{10} ~M_{\ast}/M_{\odot}>10$ with the only non-detection occurring in a PSB with $\rm log_{10} ~M_{\ast}/M_{\odot}<10$. This sample was measured to have $8.5<\rm log_{10} ~M_{H_{2}}/M_{\odot}<9.6$ and the significant quantity of observed molecular gas resulted in the conclusion that the global molecular gas reservoir is not consumed or expelled through the PSB phase. By selecting galaxies with strong Balmer absorption and shocked ionized gas emission inconsistent with star formation, \citet{alataloSHOCKEDPOSTSTARBURSTGALAXY2016a} formed a sample of shocked post-starburst (SPOGs) galaxies and obtained 52 global molecular gas measurements, detecting 90 per cent of SPOGs. Only three out of 52 galaxies in the SPOGS sample have $\rm log_{10} ~M_{\ast}/M_{\odot}<10$, and they find significant molecular gas reservoirs with gas fractions larger than even normal SF galaxies, concluding that in SPOGs, the galaxies quench despite large molecular gas content. \citet{yesuf_2017} obtained molecular gas measurements for 22 BPT selected Seyfert AGN PSBs and found, in contrast to the other previous studies, molecular gas in only 27 per cent of their PSBs. 

By selecting a mass and redshift complete sample of both E+A and PCA PSB galaxies down to a matched sensitivity threshold of a suitable comparison sample in xCOLD GASS, we have shown that the evolution of gas in PSBs is diverse, stellar mass dependent, and may proceed via multiple routes. Specifically, from Figure \ref{fig:delta-delta}, we can see that there exists a number of different PSB populations present in our sample. For galaxies with $\rm log_{10} ~M_{\ast}/M_{\odot}<10$ (blue), roughly half are gas-normal in one or both gas phases. Conversely, the other half of $\rm log_{10} ~M_{\ast}/M_{\odot}<10$ PSBs are depleted, especially in molecular gas. As we have shown for the low mass galaxies via stacking, the typical PSB not detected in CO in this regime can host gas fractions that are sub-0.5 per cent. When $\rm log_{10} ~M_{\ast}/M_{\odot}>10$ (purple and red), Figure \ref{fig:delta-delta} shows that the majority of these galaxies exist in the intermediate regime between the SF and quenched loci and are more uniformly distributed from one locus to the other. There may be a number of distinct physical processes at play potentially dependent on stellar mass.

There has been considerable work done to disentangle the various evolutionary pathways of PSB galaxies in recent years (e.g., \citeauthor{davis_2019} \citeyear{davis_2019}; \citeauthor{chen_2019} \citeyear{chen_2019};
\citeauthor{pawlikOriginsPoststarburstGalaxies2018} \citeyear{pawlikOriginsPoststarburstGalaxies2018};
\citeauthor{pawlikDiverseEvolutionaryPathways2019} \citeyear{pawlikDiverseEvolutionaryPathways2019};
\citeauthor{cheng_2024} \citeyear{cheng_2024};
\citeauthor{nielson_2025} \citeyear{nielson_2025};
\citeauthor{leung_2025} \citeyear{leung_2025}). Simulation work by \citet{pawlikDiverseEvolutionaryPathways2019} identifies four potential evolutionary pathways for PSBs: (1) blue$\rightarrow$red, where a merger results in a terminal quenching of SFR; (2) a blue cycle, where residual gas after the post-starburst phase allows for a rejuvenation of star formation; (3) a red cycle, where SFR enhancements in a quenched galaxy are due to the delivery of gas from a wet minor merger which is promptly exhausted; and (4) truncation, where the SFR halts after a fast exhaustion of the gas reservoir. Observationally, this is corroborated by \citet{pawlikOriginsPoststarburstGalaxies2018} where low-mass PSBs may return back to the main sequence given sufficient gas content, whereas higher mass PSBs are more likely to monotonically transition from SF to quenched given a merger-driven starburst. Considering the gas properties of EMBERS PSBs, we see evidence of the potential for the rejuvenation of SFR given the considerable molecular and atomic gas reservoirs seen in the gas-rich population of $\rm log_{10} ~M_{\ast}/M_{\odot}<10$ PSBs. Similarly, the gas-poor $\rm log_{10} ~M_{\ast}/M_{\odot}<10$ PSBs may be due to truncation caused by gas reservoir exhaustion. Many of the $\rm log_{10} ~M_{\ast}/M_{\odot}>10$ PSBs may be in various stages of a blue to red transition mode. Thus, we may be seeing signs of a bimodality in gas properties present in the lower stellar mass regime, but also a difference in quenching mechanisms across stellar masses. The fraction of PSBs that are caused by mergers has been shown to strongly increase with stellar mass \citep{ellison_psbs_pm_2024} which falls below 25 per cent at $\rm log_{10} ~M_{\ast}/M_{\odot} \sim 10$, and so gas exhaustion as a cause for the rapid truncation of star formation in low mass PSBs may help in reconciling the low merger fraction.

The molecular gas reservoirs present in PSBs have been predicted to exhaust with the time since the most recent starburst in both simulations (\citeauthor{davis_2019} \citeyear{davis_2019}) and observations (e.g., \citeauthor{french_2018} \citeyear{french_2018}; \citeauthor{bezanson_2022} \citeyear{bezanson_2022}; \citeauthor{frenchStateMolecularGas2023c} \citeyear{frenchStateMolecularGas2023c}). The bulk of these studies base their samples off of galaxies with $\rm log_{10} ~M_{\ast}/M_{\odot}>10$, with mostly small samples, and so this coherent loss of gas may be related to a more monotonic migration from blue to red. \citetalias{ellisonLowRedshiftPoststarburst2025a} showed that the atomic gas in EMBERS PSBs does \textit{not} deplete with time since burst which may be the result of the inclusion of low stellar mass PSBs that are quenching via different means. We leave a detailed analysis of the molecular gas properties and how they change with the burst parameters to a future work such that we can determine if there is a mass dependence on how PSBs evolve and what their star formation future holds.

Regardless of the end state of the PSBs, it is still germane to ask which gas phase is responsible for the rapid quenching of SFR that we do see. Referring back to Figure \ref{fig:delta-delta}, if molecular gas is depleted or evacuated first, a galaxy would travel first down and then to the left. Conversely, if it is the atomic phase that dissipates first, unable to fuel the molecular gas reservoir, a galaxy would travel first left and then down. Coherent loss of gas due to consumption or evacuation of both phases would result in a galaxy traveling from the SF locus to the quenched region. There exists a dearth of PSBs along the connecting line between the loci in Figure \ref{fig:delta-delta}, meaning that this coherent approach to gas depletion is not favoured by our PSBs. Most of our PSBs have elevated $\Delta \rm M_{H_2}$ for a given $\Delta \rm M_{H\text{\sc{i}}}$ (i.e., above the 1-to-1 line) which we can interpret as these PSBs depleting in atomic gas relative to molecular gas. Most of the galaxies in this case are of lower stellar mass. There also exists a population of PSBs with depleted molecular gas reservoirs relative to their atomic gas (i.e., below the 1-to-1 line) which is where 75 per cent of the highest mass PSBs exist. A more detailed accounting of the star formation rates and histories is needed to determine the relative role of each gas phase in EMBERS PSBs, which we will present in a follow-up paper.


\section{Conclusions}
\label{sec:conclusions}
Studying post-starburst galaxies allows us to directly investigate the properties of galaxies that are in the process of quenching their star formation. In this work, we obtained new measurements of molecular gas in PSBs with CO(1--0) observations at the IRAM 30m telescope. These observations were combined with the comparable observations in the available literature to form a homogeneous sample of 61 galaxies spanning stellar masses of log$_{10}$~M$_{\ast}$/M$_{\odot}$$>9.56$ and a redshift range of $0.01<z<0.0362$. Of these galaxies, 58 also have supporting atomic gas observations from \citetalias{ellisonLowRedshiftPoststarburst2025a}, allowing us to perform the first complete census of multi-phase gas in PSBs which we call EMBERS, or the Ensemble of Multiphase Baryons Evolving in Rapidly-quenching Systems. We are also able to utilize a rigorous control sample of equivalent survey design, in the forms of xGASS and xCOLD GASS, to place EMBERS PSBs in the context of a representative set of galaxies. Our findings are as follows:

\begin{itemize}
    \item \textbf{PSBs host molecular gas reservoirs in the majority of galaxies}. We detect CO emission at the $5\sigma$ level in 34/61 galaxies with inferred molecular gas masses from detections ranging from $8.2<\rm log_{10} ~M_{H_{2}}/M_{\odot}<10.3$ corresponding to molecular gas fractions ranging from 2 per cent to almost 250 per cent (see Figure \ref{fig:gass distributions}). One post-starburst, EMB85, hosts a larger molecular gas reservoir than its stellar mass component. In the higher stellar mass range, the molecular gas reservoirs we observe in both the total gas mass and gas fraction relative to stellar mass are consistent with the previous literature.\\ 
    \item \textbf{The typical molecular gas mass in PSBs is between $\sim$0.3 and 0.6 dex less than their star forming progenitors.} At lower stellar masses ($ \rm log_{10} ~M_{\ast}/M_{\odot}<10.4$) they are also $\sim 0.2$ dex deficient in molecular gas compared to \textit{all} normal galaxies, a deficit which disappears in the higher mass regime (see Figure \ref{fig:fractions}). Additionally, the rate of $5\sigma$ detections is consistently less than the entire xCOLD GASS sample for  $\rm log_{10} ~M_{\ast}/M_{\odot}<10.4$ and increases with stellar mass for our PSBs.\\
    \item \textbf{The molecular-to-atomic gas ratio in PSBs is not systematically lower than in regular galaxies, as might be expected if the conversion from H$\text{\sc{i}}$ to H$_2$ is the cause of quenching.} The typical value of $\rm R_{mol}=log_{10}~M_{H_2}/M_{H\text{\sc{i}}}$ in EMBERS PSBs is $~0.2$ dex \textit{larger} compared to the entire xGASS/xCOLD GASS sample. When compared to just star forming galaxies in the same mass regime, the molecular-to-atomic gas ratio in EMBERS PSBs are consistent with normal SF galaxies (see Figure \ref{fig:molecular-atomic}). One PSB, EMB12, has $10$x more molecular gas than atomic gas.\\
    \item \textbf{Although median gas fractions in PSBs are depleted compared with SF controls, individual galaxies show a diversity of $\Delta$M$_{\rm H_2}$}. Matching each of our PSBs to a bespoke xCOLD GASS control sample allows us to compute each galaxy's relative H$_2$-richness (see Figure \ref{fig:deltas}). 32/61 PSBs have gas mass offsets below the $1\sigma$ spread of xCOLD GASS and are thus significantly gas poor with deficiencies ranging from $\sim 0.5 - 1.3$ dex. That being said, many PSBs are molecular gas-normal with an additional $\sim$10 galaxies showing signs of enhanced gas fractions compared to SF galaxies from xCOLD GASS. The (qualitative) gas deficiencies that we observe are robust against choices of $\rm \alpha_{CO}$ (see Section \ref{sec:alpha} and Figure \ref{fig:change in alphas}). \\
    \item \textbf{The gas reservoirs, in both the atomic and molecular phase, are intermediate in mass between gas-rich, star-forming galaxies and gas-poor, quenched galaxies.} Many of our PSBs occupy a regime in $\Delta \rm M_{H\text{\sc{i}}}$ and $\Delta\rm M_{H_2}$ in between SF and quenched xGASS/xCOLD GASS galaxies largely not occupied by the comparison sample, confirming their identification of having rapidly quenched in SFR (see Figure \ref{fig:delta-delta}). Further analysis of the entire star formation histories of these PSB galaxies is needed to disentangle the relative importance of both gas phases and the relationship between gas depletion and current star formation. \\
    \item \textbf{Multiple physical processes are likely required to explain the origin of post-starburst galaxies}. We have identified a population of low mass PSBs that have molecular gas fractions that are sub-0.5 per cent (with low H$\text{\sc{i}}$ reservoirs) and so the rapid reduction in SFR seen in these PSBs may be due to total gas depletion (see Figure \ref{fig:result of stacks}). Conversely, there are many PSBs that are gas-normal to gas-rich in either H$\text{\sc{i}}$ or $\rm H_2$ (or both) such that major disruptions to the ISM are likely responsible for the quenching of star formation in these cases (see Figure \ref{fig:delta-delta}). The considerable combined gas reservoirs found in some of our PSBs could potentially allow for the rejuvenation of star formation in the future.
\end{itemize}

In future work, we will present CO(2--1) observations of the same sample of PSBs and investigate how the ratio of CO(2--1)/CO(1--0) behaves in the rapidly quenched regime. Additionally, we will extend the analysis presented here by fitting the star formation histories of the EMBERS sample, which will allow us to constrain the current rates of star formation and properties of the most recent starburst. Combined with the atomic gas phase measurements from \citetalias{ellisonLowRedshiftPoststarburst2025a} and the CO(1--0) observations presented here, we will examine the star formation efficiencies of EMBERS PSBs, determine the relative importance of each gas phase in dictating the nature of quenching in our PSBs, and evaluate how their gas properties evolve with time since burst (Rasmussen et al., in prep.).

\section*{Acknowledgments}
\label{acknowledgements}

We respectfully acknowledge the L\textschwa\textvbaraccent {k}$^{\rm w}$\textschwa\ng{}\textschwa n Peoples on whose traditional territory the University of Victoria stands and the Songhees, Esquimalt and $\underline{\text{W}}\acute{\text{S}}$ANE$\acute{\text{C}}$ peoples whose relationships with the land continue to this day. As we explore the shared sky, we acknowledge our responsibilities to honour those who were here before us, and their continuing relationships to these lands. We strive for respectful relationships and partnerships with all the peoples of these lands as we move forward together towards reconciliation and decolonization. 

{We thank the referee for their input and constructive feedback which improved this manuscript.} BR would like to thank Dave Patton, Simon Smith, and Aydan McKay for their scientific discussions that improved the quality of this work. BR would also like to thank the IRAM 30m staff for the help provided during the observations of this project. SW gratefully acknowledges the support from the Natural Sciences and Engineering Council of Canada (NSERC) as part of their graduate fellowship program. SLE gratefully acknowledges the receipt of NSERC Discovery Grants. VW acknowledges Science and Technologies Facilities Council (STFC) grant ST/Y00275X/1, and Leverhulme Research Fellowship RF-2024-589/4. JW thanks the support of research grants from
Ministry of Science and Technology of the People’s Republic of China (NO. 2022YFA1602902),
National Science Foundation of China (NO. 12233001), and the China Manned Space Project (No. CMS-CSST-2025-A08). 

This work is based on observations carried out under projects numbered 144-23, 077-24, 180-24, and 077-25 with the IRAM 30m telescope. IRAM is supported by INSU/CNRS (France), MPG (Germany) and IGN (Spain). This work also made use of data from FAST (Five-hundred-metre Aperture Spherical radio Telescope) (\url{https://cstr.cn/31116.02.FAST}). FAST is a Chinese national mega-science
facility, operated by the National Astronomical Observatories, Chinese Academy of Sciences.

Funding for the SDSS and SDSS-II has been provided by the Alfred P. Sloan Foundation, the Participating Institutions, the National Science Foundation, the U.S. Department of Energy, the National Aeronautics and Space Administration, the Japanese Monbukagakusho, the Max Planck Society, and the Higher Education Funding Council for England. The SDSS Web Site is \url{http://www.sdss.org/.}

The SDSS is managed by the Astrophysical Research Consortium for the Participating Institutions. The Participating Institutions are the American Museum of Natural History, Astrophysical Institute Potsdam, University of Basel, University of Cambridge, Case Western Reserve University, University of Chicago, Drexel University, Fermilab, the Institute for Advanced Study, the Japan Participation Group, Johns Hopkins University, the Joint Institute for Nuclear Astrophysics, the Kavli Institute for Particle Astrophysics and Cosmology, the Korean Scientist Group, the Chinese Academy of Sciences (LAMOST), Los Alamos National Laboratory, the Max-Planck-Institute for Astronomy (MPIA), the Max-Planck-Institute for Astrophysics (MPA), New Mexico State University, Ohio State University, University of Pittsburgh, University of Portsmouth, Princeton University, the United States Naval Observatory, and the University of Washington.

The Legacy Surveys consist of three individual and complementary projects: the Dark Energy Camera Legacy Survey (DECaLS; Proposal ID \#2014B-0404; PIs: David Schlegel and Arjun Dey), the Beijing-Arizona Sky Survey (BASS; NOAO Prop. ID \#2015A-0801; PIs: Zhou Xu and Xiaohui Fan), and the Mayall z-band Legacy Survey (MzLS; Prop. ID \#2016A-0453; PI: Arjun Dey). DECaLS, BASS and MzLS together include data obtained, respectively, at the Blanco telescope, Cerro Tololo Inter-American Observatory, NSF’s NOIRLab; the Bok telescope, Steward Observatory, University of Arizona; and the Mayall telescope, Kitt Peak National Observatory, NOIRLab. Pipeline processing and analyses of the data were supported by NOIRLab and the Lawrence Berkeley National Laboratory (LBNL). The Legacy Surveys project is honored to be permitted to conduct astronomical research on Iolkam Du’ag (Kitt Peak), a mountain with particular significance to the Tohono O’odham Nation.

NOIRLab is operated by the Association of Universities for Research in Astronomy (AURA) under a cooperative agreement with the National Science Foundation. LBNL is managed by the Regents of the University of California under contract to the U.S. Department of Energy.

This project used data obtained with the Dark Energy Camera (DECam), which was constructed by the Dark Energy Survey (DES) collaboration. Funding for the DES Projects has been provided by the U.S. Department of Energy, the U.S. National Science Foundation, the Ministry of Science and Education of Spain, the Science and Technology Facilities Council of the United Kingdom, the Higher Education Funding Council for England, the National Center for Supercomputing Applications at the University of Illinois at Urbana-Champaign, the Kavli Institute of Cosmological Physics at the University of Chicago, Center for Cosmology and Astro-Particle Physics at the Ohio State University, the Mitchell Institute for Fundamental Physics and Astronomy at Texas A\&M University, Financiadora de Estudos e Projetos, Fundacao Carlos Chagas Filho de Amparo, Financiadora de Estudos e Projetos, Fundacao Carlos Chagas Filho de Amparo a Pesquisa do Estado do Rio de Janeiro, Conselho Nacional de Desenvolvimento Cientifico e Tecnologico and the Ministerio da Ciencia, Tecnologia e Inovacao, the Deutsche Forschungsgemeinschaft and the Collaborating Institutions in the Dark Energy Survey. The Collaborating Institutions are Argonne National Laboratory, the University of California at Santa Cruz, the University of Cambridge, Centro de Investigaciones Energeticas, Medioambientales y Tecnologicas-Madrid, the University of Chicago, University College London, the DES-Brazil Consortium, the University of Edinburgh, the Eidgenossische Technische Hochschule (ETH) Zurich, Fermi National Accelerator Laboratory, the University of Illinois at Urbana-Champaign, the Institut de Ciencies de l’Espai (IEEC/CSIC), the Institut de Fisica d’Altes Energies, Lawrence Berkeley National Laboratory, the Ludwig Maximilians Universitat Munchen and the associated Excellence Cluster Universe, the University of Michigan, NSF’s NOIRLab, the University of Nottingham, the Ohio State University, the University of Pennsylvania, the University of Portsmouth, SLAC National Accelerator Laboratory, Stanford University, the University of Sussex, and Texas A\&M University.

BASS is a key project of the Telescope Access Program (TAP), which has been funded by the National Astronomical Observatories of China, the Chinese Academy of Sciences (the Strategic Priority Research Program “The Emergence of Cosmological Structures” Grant \# XDB09000000), and the Special Fund for Astronomy from the Ministry of Finance. The BASS is also supported by the External Cooperation Program of Chinese Academy of Sciences (Grant \# 114A11KYSB20160057), and Chinese National Natural Science Foundation (Grant \# 12120101003, \# 11433005).

The Legacy Survey team makes use of data products from the Near-Earth Object Wide-field Infrared Survey Explorer (NEOWISE), which is a project of the Jet Propulsion Laboratory/California Institute of Technology. NEOWISE is funded by the National Aeronautics and Space Administration.

The Legacy Surveys imaging of the DESI footprint is supported by the Director, Office of Science, Office of High Energy Physics of the U.S. Department of Energy under Contract No. DE-AC02-05CH1123, by the National Energy Research Scientific Computing Center, a DOE Office of Science User Facility under the same contract; and by the U.S. National Science Foundation, Division of Astronomical Sciences under Contract No. AST-0950945 to NOAO.
\section*{Data Availability}

The CO(1--0) data presented in this paper can be made available upon reasonable request of the first author.



\bibliographystyle{mnras}
\bibliography{bibliography}

\bsp	
\appendix

\suppressfloats[t]





\section{Additional spectra and stacks }
\label{app:additional spectra}

\begin{figure}
  \centering
  \includegraphics[width=0.35\textwidth]{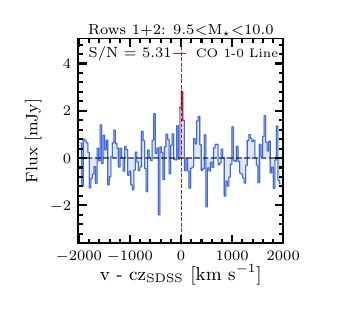}\hspace{-0.03\textwidth}

  \caption{Results of stacking all non-detected CO spectra for galaxies with log$_{10}$ M$_{\ast}$/M$_{\odot}$$<10.0$. The stacked spectrum includes equivalent integration time of over 100 hours with the IRAM 30m and results in a detection at the $5\sigma$ level corresponding to an inferred gas fraction of $<0.5$ per cent.}

  \label{fig:stacks low mass}
\end{figure}

\begin{figure*}
  \centering
  \includegraphics[width=0.5\textwidth]{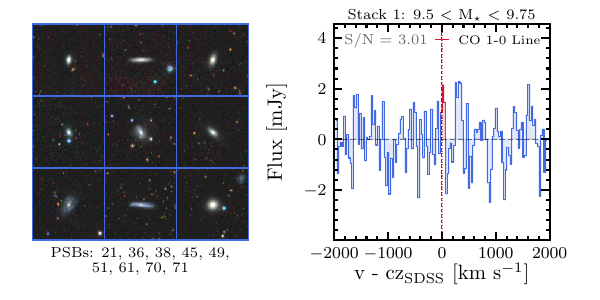}\hspace{-0.01\textwidth}
  \includegraphics[width=0.5\textwidth]{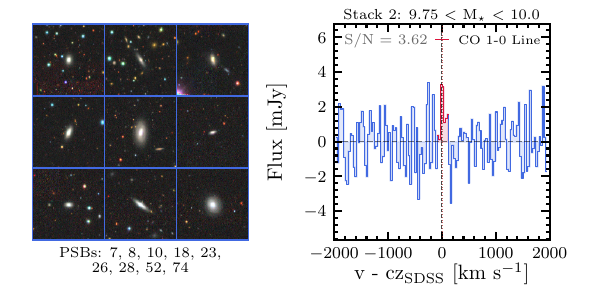}\\[0.005\textwidth]
  \includegraphics[width=0.5\textwidth]{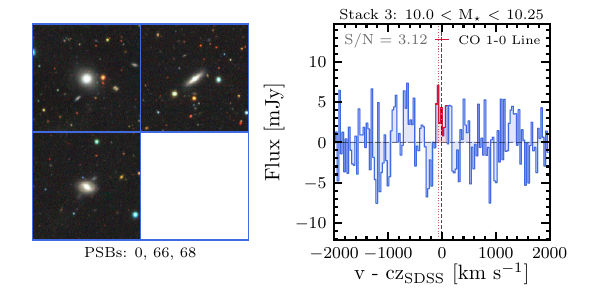}\hspace{-0.01\textwidth}
  \includegraphics[width=0.5\textwidth]{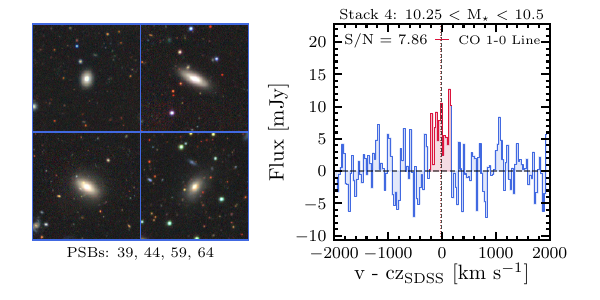}\\[-0.01\textwidth]

  \caption{Results of spectral stacking of non-detected CO spectra in bins log$_{10}$ M$_{\ast}$/M$_{\odot}$$=0.25$. Shown to the left of each spectral stack are the DECaLs images of the individual galaxies involved in each stack and their corresponding PSB sample identifiers. Only the highest stellar mass bin is detected at the $5\sigma$ level.}

  \label{fig:stacks all}
\end{figure*}


\begin{figure*}
    \centering
    \newcommand{\vertspace}{-0.1cm}
    \includegraphics[width=0.49\textwidth]{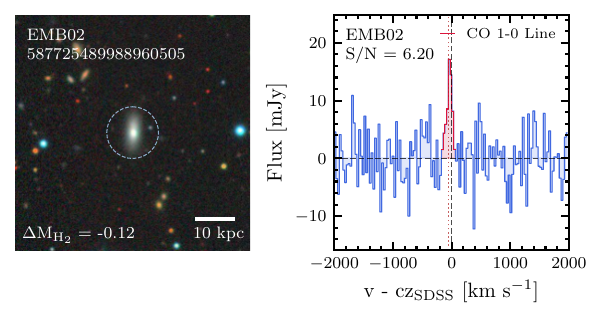}
    \includegraphics[width=0.49\textwidth]{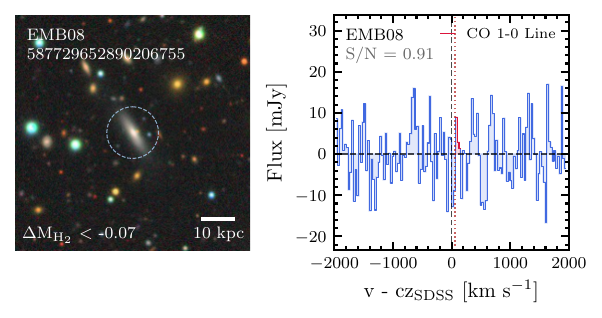}\\ \vspace{15pt}
    \includegraphics[width=0.49\textwidth]{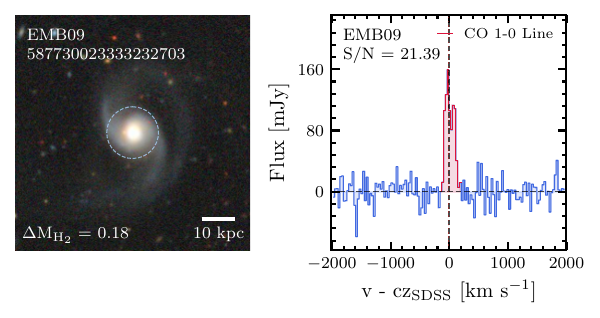}
    \includegraphics[width=0.49\textwidth]{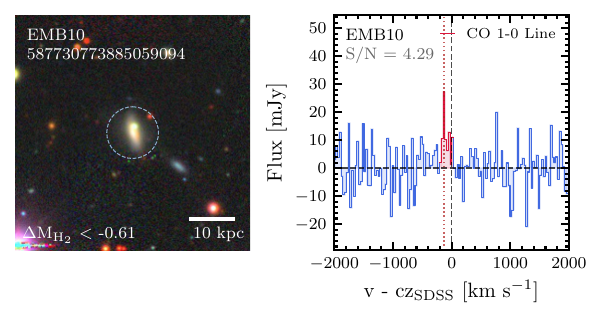}\\ \vspace{15pt}
    \caption{Figure~\ref{fig:psb_spectra_1} continued.}
    \label{fig:psb_spectra_2}
\end{figure*}

\begin{figure*}
    \centering
    \newcommand{\vertspace}{-0.1cm}
    \includegraphics[width=0.49\textwidth]{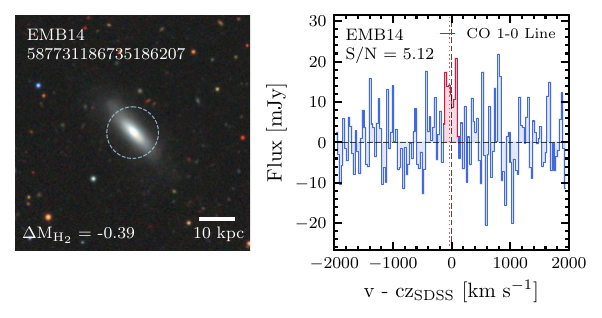}
    \includegraphics[width=0.49\textwidth]{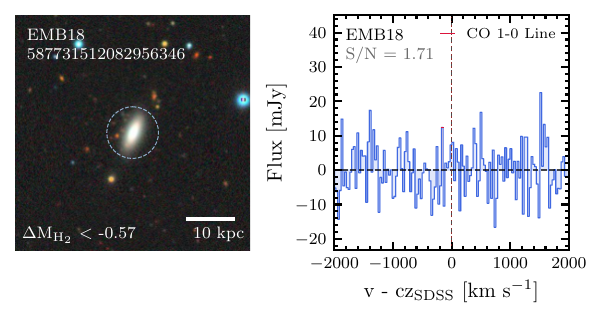}\\ \vspace{15pt}
    \includegraphics[width=0.49\textwidth]{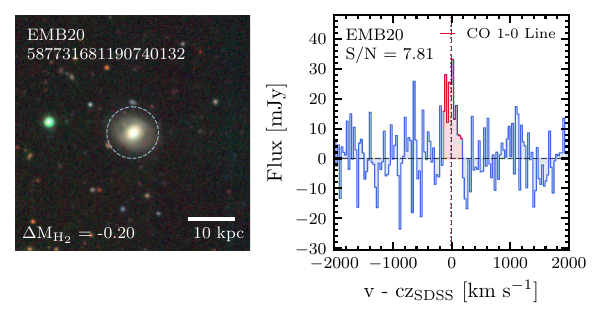}
    \includegraphics[width=0.49\textwidth]{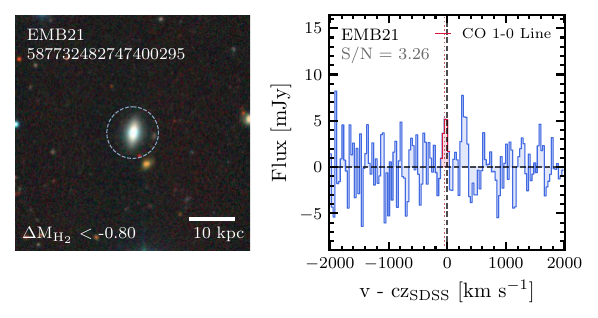}\\ \vspace{15pt}
    \includegraphics[width=0.49\textwidth]{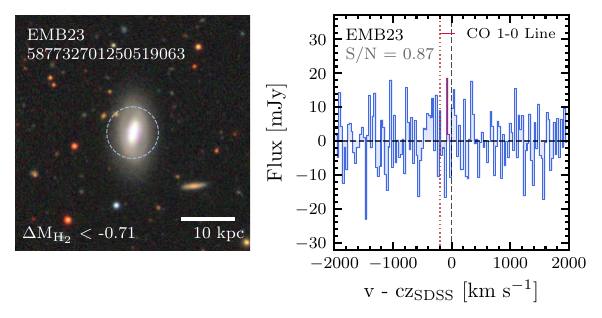}
    \includegraphics[width=0.49\textwidth]{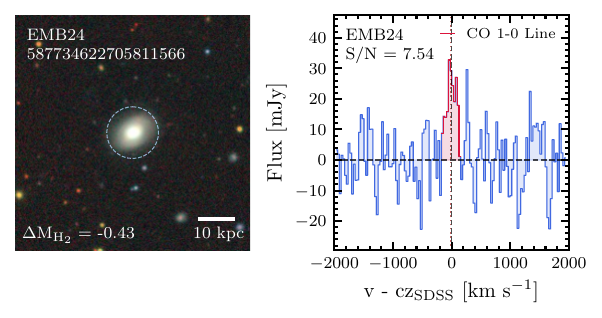}\\ \vspace{15pt}
    \includegraphics[width=0.49\textwidth]{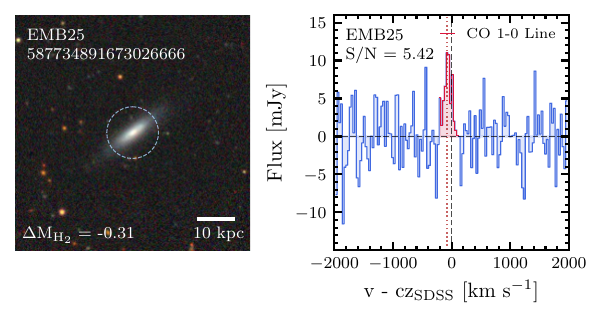}
    \includegraphics[width=0.49\textwidth]{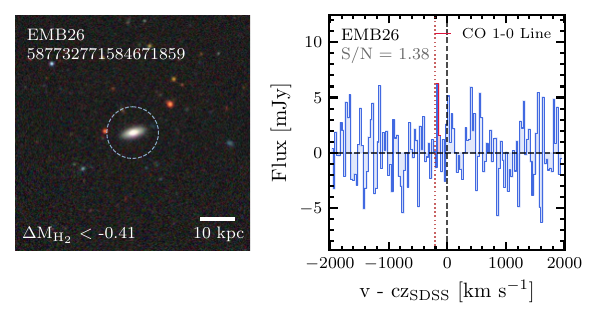}
    \caption{Figure~\ref{fig:psb_spectra_1} continued.}
    \label{fig:psb_spectra_3}
\end{figure*}

\begin{figure*}
    \centering
    \newcommand{\vertspace}{-0.1cm}
    \includegraphics[width=0.49\textwidth]{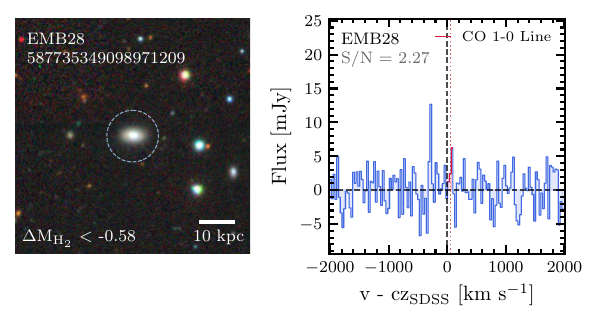}
    \includegraphics[width=0.49\textwidth]{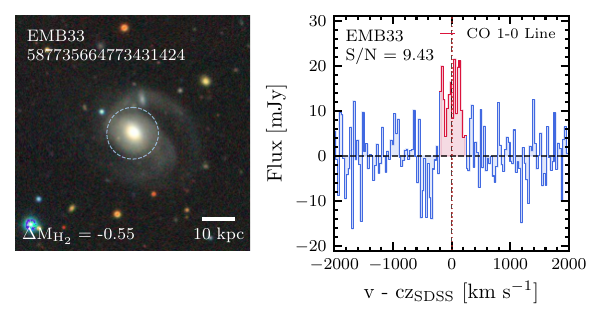}\\ \vspace{15pt}
    \includegraphics[width=0.49\textwidth]{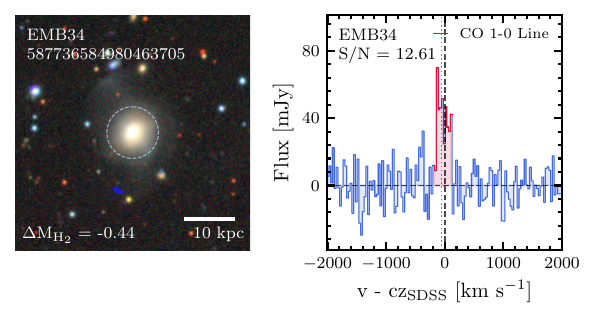}
    \includegraphics[width=0.49\textwidth]{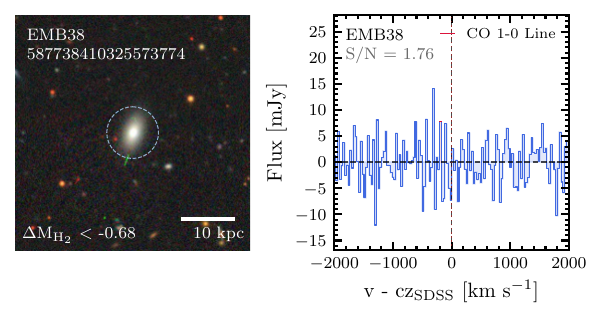}\\ \vspace{15pt}
    \includegraphics[width=0.49\textwidth]{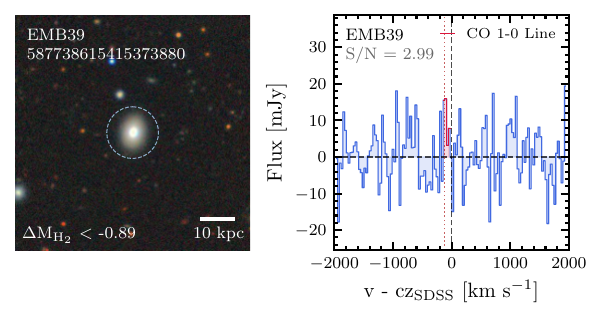}
    \includegraphics[width=0.49\textwidth]{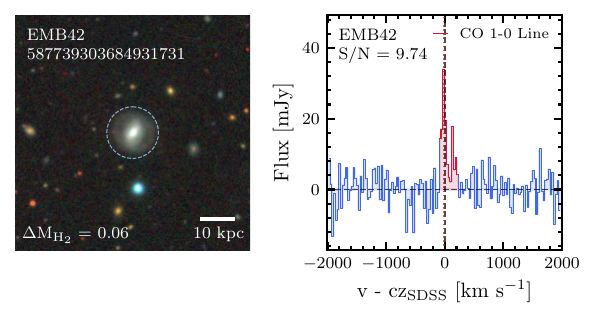}\\ \vspace{15pt}
    \includegraphics[width=0.49\textwidth]{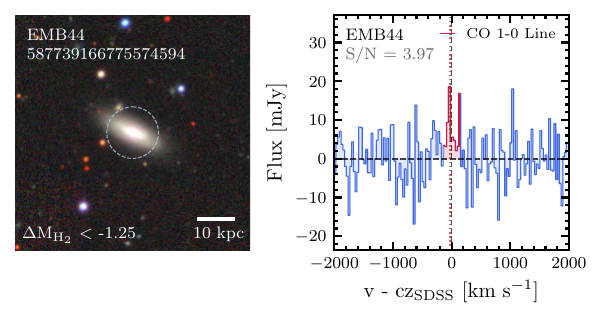}
    \includegraphics[width=0.49\textwidth]{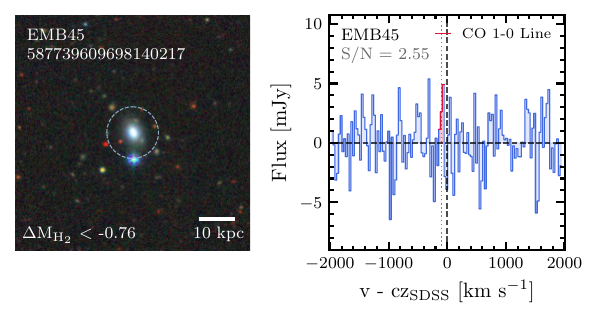}
    \caption{Figure~\ref{fig:psb_spectra_1} continued.}
    \label{fig:psb_spectra_4}
\end{figure*}

\begin{figure*}
    \centering
    \newcommand{\vertspace}{-0.1cm}
    \includegraphics[width=0.49\textwidth]{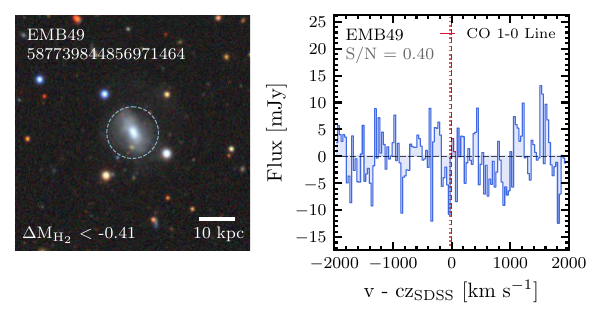}
    \includegraphics[width=0.49\textwidth]{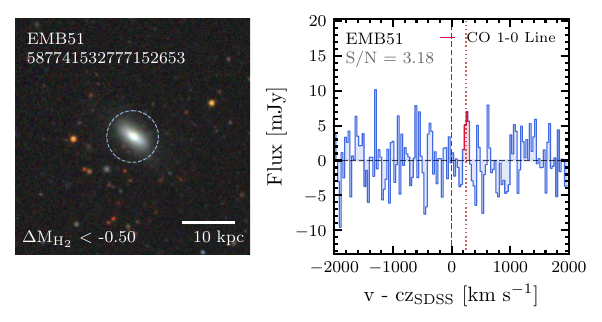}\\ \vspace{15pt}
    \includegraphics[width=0.49\textwidth]{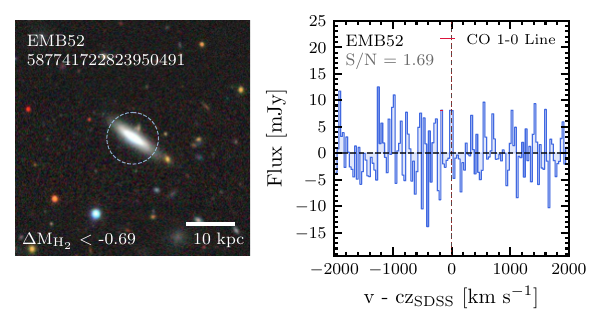}
    \includegraphics[width=0.49\textwidth]{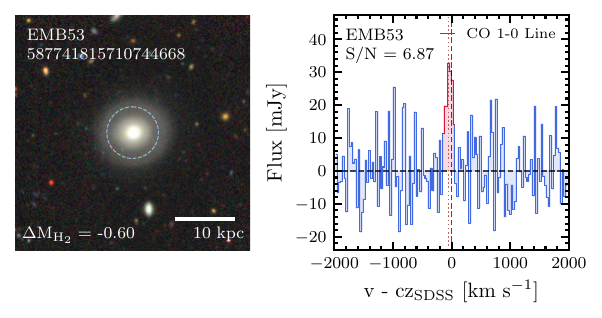}\\ \vspace{15pt}
    \includegraphics[width=0.49\textwidth]{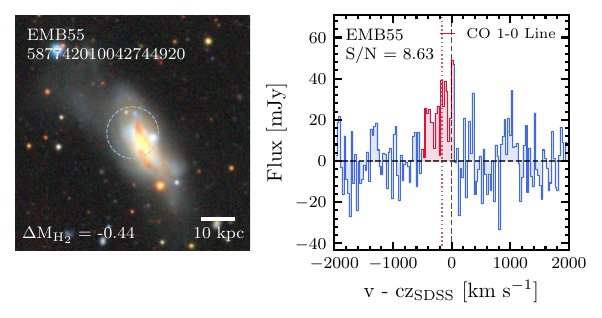}
    \includegraphics[width=0.49\textwidth]{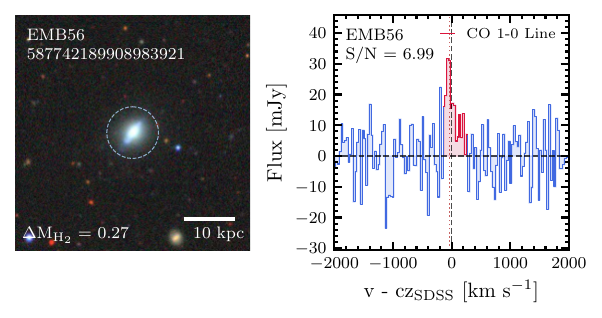}\\ \vspace{15pt}
    \includegraphics[width=0.49\textwidth]{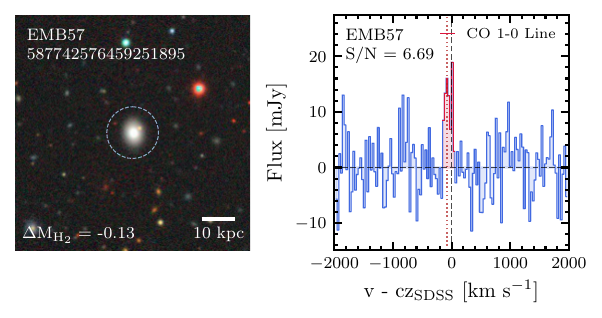}
    \includegraphics[width=0.49\textwidth]{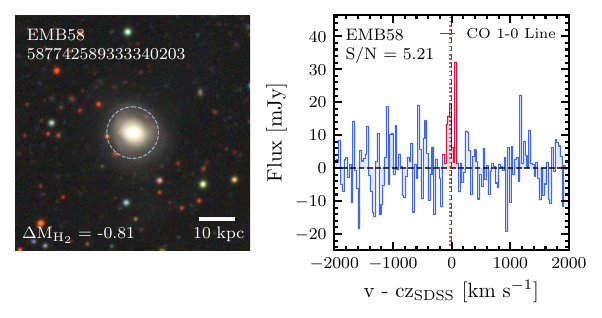}
    \caption{Figure~\ref{fig:psb_spectra_1} continued.}
    \label{fig:psb_spectra_5}
\end{figure*}

\begin{figure*}
    \centering
    \newcommand{\vertspace}{-0.1cm}
    \includegraphics[width=0.49\textwidth]{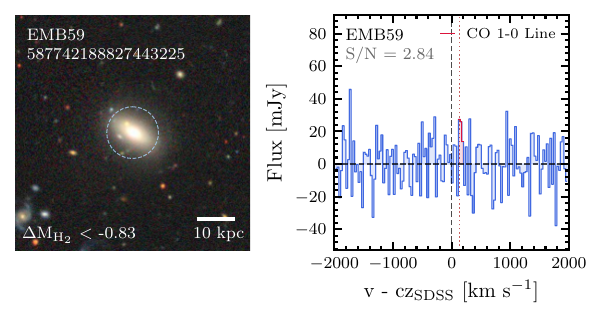}
    \includegraphics[width=0.49\textwidth]{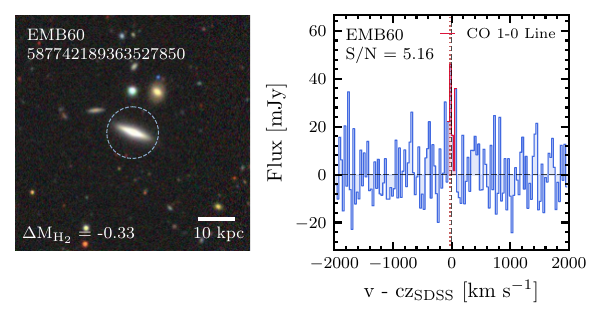}\\ \vspace{15pt}
    \includegraphics[width=0.49\textwidth]{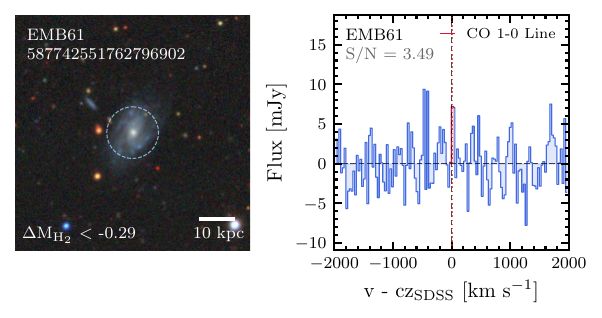}
    \includegraphics[width=0.49\textwidth]{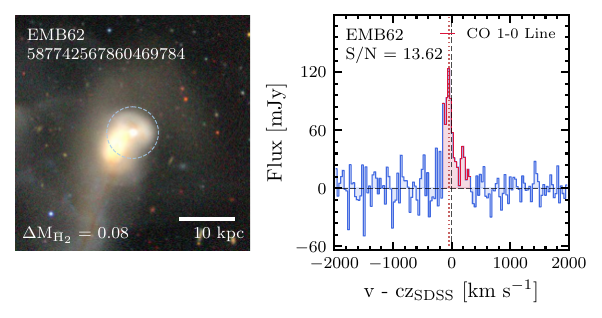}\\ \vspace{15pt}
    \includegraphics[width=0.49\textwidth]{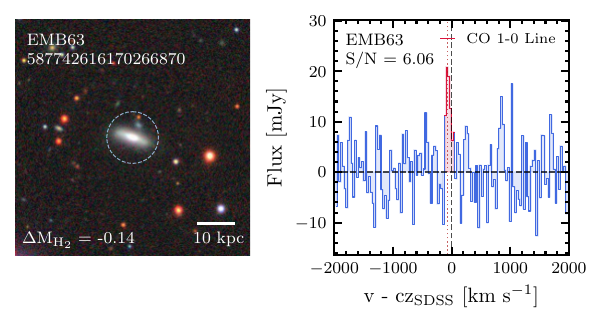}
    \includegraphics[width=0.49\textwidth]{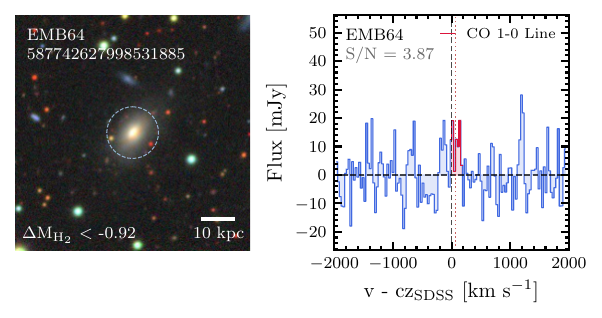}\\ \vspace{15pt}
    \includegraphics[width=0.49\textwidth]{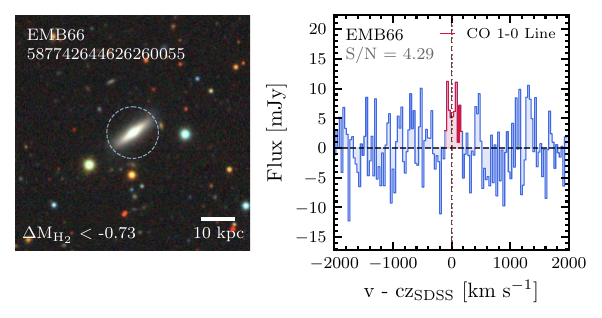}
    \includegraphics[width=0.49\textwidth]{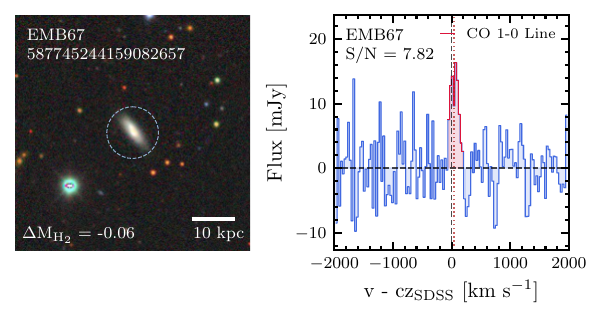}
    \caption{Figure~\ref{fig:psb_spectra_1} continued.}
    \label{fig:psb_spectra_6}
\end{figure*}

\begin{figure*}
    \centering
    \newcommand{\vertspace}{-0.1cm}
    \includegraphics[width=0.49\textwidth]{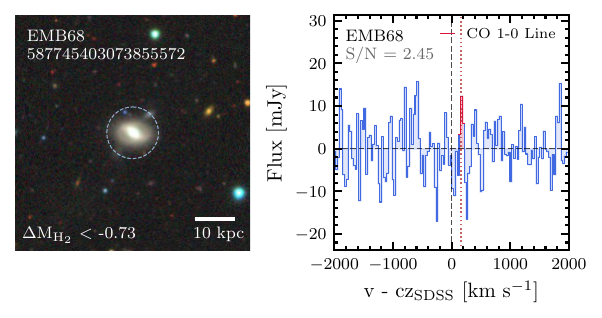}
    \includegraphics[width=0.49\textwidth]{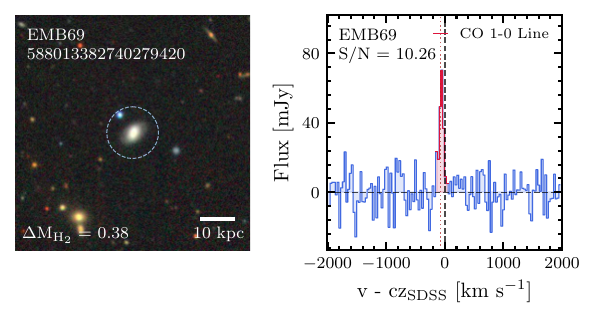}\\ \vspace{15pt}
    \includegraphics[width=0.49\textwidth]{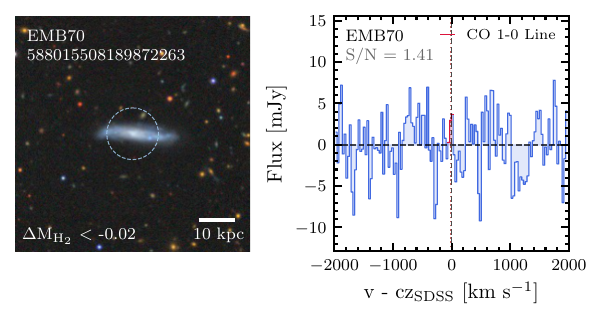}
    \includegraphics[width=0.49\textwidth]{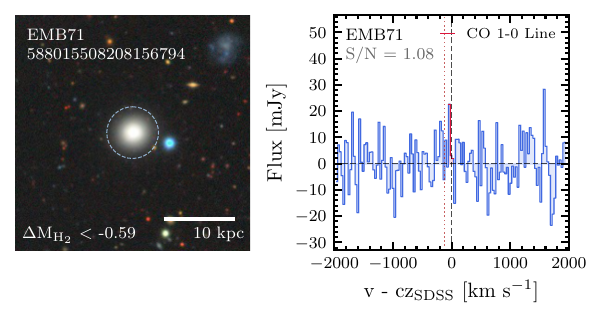}\\ \vspace{15pt}
    \includegraphics[width=0.49\textwidth]{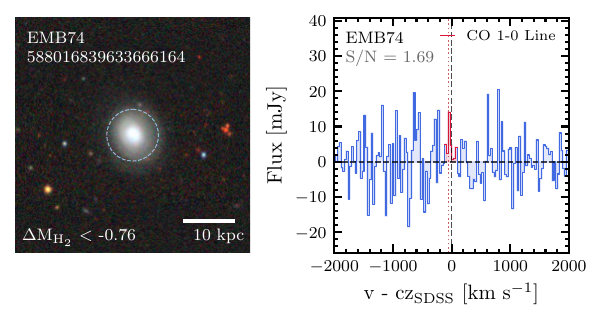}
    \includegraphics[width=0.49\textwidth]{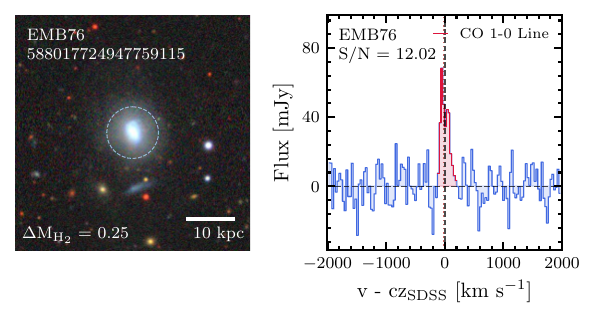}\\ \vspace{15pt}
    \includegraphics[width=0.49\textwidth]{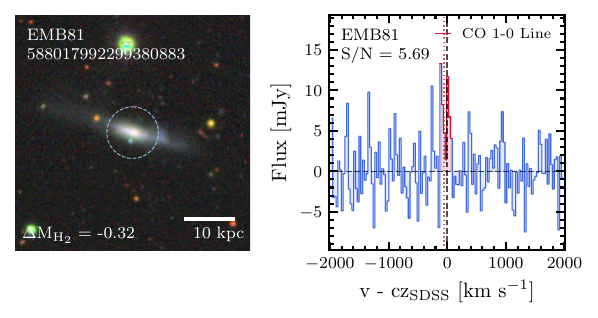}
    \includegraphics[width=0.49\textwidth]{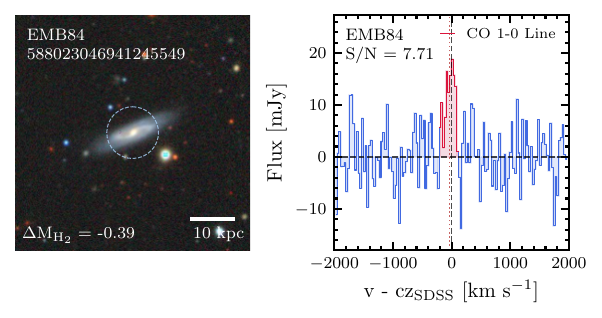}
    \caption{Figure~\ref{fig:psb_spectra_1} continued.}
    \label{fig:psb_spectra_7}
\end{figure*}


\label{lastpage}
\end{document}